%% file: root.tex
\documentclass[twocolumn]{autart}  
\usepackage[T1]{fontenc}

\usepackage{amsmath,amssymb,mathtools}
\usepackage[shortlabels]{enumitem}
\usepackage{graphicx}
\usepackage{xcolor}
\usepackage[flushleft]{threeparttable}

\usepackage[absolute,overlay]{textpos}
\usepackage[colorlinks=true,urlcolor=teal]{hyperref}

\begin{document}

\begin{frontmatter}

\title{Nonlinear moving horizon estimation for robust state and parameter estimation -- extended version\thanksref{footnoteinfo}}

\thanks[footnoteinfo]{The material in this paper was partially presented at the 2023 IFAC World Congress, July 9–14, 2023, Yokohama, Japan. Corresponding author Julian D. Schiller. Tel. +49-511-762-18902. Fax +49-511-762-4536.}

\author[]{Julian D. Schiller}\ead{schiller@irt.uni-hannover.de},
\author[]{Matthias A. M{\"u}ller}\ead{mueller@irt.uni-hannover.de}

\address{Leibniz University Hannover, Institute of Automatic Control, Hannover, Germany}
          

\begin{abstract} 
	We propose a moving horizon estimation scheme to estimate the states and the unknown constant parameters of general nonlinear uncertain discrete-time systems. The proposed framework and analysis explicitly do not involve the \textit{a priori} verification of a particular excitation condition for the parameters. Instead, we use online information about the actual excitation of the parameters at any time during operation and ensure that the regularization term in the cost function is always automatically selected appropriately. This ensures that the state and parameter estimation error is bounded for all times, even if the parameters are never (or only rarely) excited during operation. Robust exponential stability of the state and parameter estimation error emerges under an additional uniform condition on the maximum duration of insufficient excitation. The theoretical results are illustrated by a numerical example.
\end{abstract}

\end{frontmatter}

\begin{textblock*}{19cm}(2.1cm,26.8cm) 
	\footnotesize This is the extended version of the original paper published in Automatica: \href{https://doi.org/10.1016/j.automatica.2025.112790}{DOI 10.1016/j.automatica.2025.112790}.
\end{textblock*}

\section{Introduction}\label{sec:intro}
Robust state estimation for nonlinear systems subject to noise is a problem of high practical relevance.
However, if the underlying model is also uncertain and cannot accurately capture the real system behavior, the estimation error may even become unstable if this is not taken into account in the observer design. We address this issue by developing a robust moving horizon estimation (MHE) scheme for simultaneously estimating the unknown states and (constant) parameters of a nonlinear system.

MHE is an optimization-based technique and therefore naturally applicable to nonlinear, potentially constrained systems.
Strong robust stability properties of MHE emerge under a mild detectability condition (namely, incremental input/output-to-state stability, i-IOSS), see, e.g., \cite{Allan2021a,Knuefer2023,Hu2023,Schiller2023c}.
These guarantees, however, rely on an exact model of the real system and are therefore not necessarily valid in the case of (parametric) model uncertainties.
To address this problem, a $\min$-$\max$ MHE scheme was proposed earlier in \cite{Alessandri2012}, where at each time step a least-squares cost function is minimized for the worst case of the model uncertainties.
Yet it is often advantageous to not only ensure robustness against model errors, but also to obtain an estimate of the uncertain parameters, since a good model is crucially required for, e.g., (high-performance) control, system monitoring, or fault detection.
In this context, an MHE scheme was proposed in \cite{Sui2011}, treating the parameters as additional states (with constant dynamics), where the temporary loss of observability (due to lack of excitation) is handled by suitable regularization and adaptive weights.
However, the robustness properties have not been analyzed, and the imposed conditions for guaranteed state and parameter convergence are not trivial to verify in practice.
In~\cite{Flayac2023}, MHE under a non-uniform observability condition is considered, which is potentially also suitable to be used for joint state and parameter estimation.
The results, however, rely on regularly persistent inputs and, in particular, no fallback strategy is provided in case a lack of excitation occurs in practice during estimation.

An alternative approach to simultaneous state and parameter estimation is provided by \textit{adaptive observers}, which have been extensively studied in the literature, see, e.g., \cite{Ioannou2012}.
Theoretical guarantees usually involve a detectability/observability condition on the system states and a persistence of excitation (PE) condition to establish parameter convergence.
Different system classes (usually neglecting disturbances) have been considered, e.g., linear time-varying (LTV) systems \cite{Ticlea2016}, Lipschitz nonlinear systems under a linear parameterization \cite{Cho1997}, nonlinearly parameterized systems \cite{Farza2009}, or systems in a certain nonlinear adaptive observer canonical form, cf., e.g., \cite{Bastin1988,Marino2001}.
Alternative approaches for systems in canonical forms can be found in, e.g., \cite{Bin2021}, where more general identifiers are used to estimate the dynamics.

Many results on state and parameter estimation rely on PE conditions that are uniform in time, which is usually restrictive and cannot be guaranteed \textit{a priori} (except for, e.g., linear systems).
To ensure practical applicability, it is essential to investigate weaker (especially non-uniform) excitation conditions.
In this context, adaptive observer designs are proposed in, e.g., \cite{Marco2022,Tomei2023}, where boundedness of the state and parameter estimation error is guaranteed without excitation, and exponential stability in the presence of PE.
Relaxed excitation conditions have recently received much attention in the context of (pure) parameter estimation of regression models.
In \cite{Efimov2019}, it was shown that weaker conditions than PE, however, generally only allow for non-uniform asymptotic stability of LTV systems (that describe the error dynamics of, e.g., simple least squares estimators under linear regression models), which is also consistent with earlier works, e.g.,~\cite{Panteley2001}.
Using the \textit{dynamic regressor extension and mixing} idea, exponential convergence could be established for linear regression models (and certain classes of nonlinear ones), merely assuming interval excitation (which is strictly weaker than uniform PE), cf.,~e.g.,~\cite{Korotina2022,Ortega2022}.

\textit{Contribution:} We propose an MHE scheme for joint state and parameter estimation for general nonlinear discrete-time systems subject to process disturbances and measurement noise.
Our arguments are based on recent MHE results \cite{Allan2021a,Knuefer2023,Schiller2023c}, where only state (but no parameter) estimation is considered.
The MHE scheme avoids a uniform PE condition and instead uses online information about the current excitation of the parameters to suitably adjust the regularization term in the cost function (Section~\ref{sec:MHE_nonPE}).
We establish a bound on the state and parameter estimation error that is valid for all times (even if the parameters are never or only rarely excited), which improves the more often sufficient excitation is present. The bound specializes to a robust global exponential stability property under an additional uniform condition on the maximum duration of insufficient excitation.
A~method to online monitor the level of excitation that is applicable to general nonlinear systems is provided in Section~\ref{sec:dIOSS}, compare also \cite{Schiller2025a} for further details.
In combination, we provide a flexible MHE scheme for robust joint state and parameter estimation with intuitive error bounds that are valid independent of the parameter excitation. The numerical example in Section~\ref{sec:example} illustrates that the proposed approach is able to efficiently compensate for phases of weak excitation and ensure reliable estimation results for all times.

In the preliminary conference version~\cite{Schiller2022a}, we have established exponential convergence of MHE using a joint i-IOSS Lyapunov function for both the states and the parameters, and we have provided a constructive approach for the computation of joint i-IOSS Lyapunov functions that is applicable to a special class of nonlinear systems. In this paper, we show that the existence of such a joint i-IOSS Lyapunov function is equivalent to the detectability of the states and a uniform PE condition of the parameters (which is restrictive).
Compared to~\cite{Schiller2022a}, we consider the much more practically relevant case where the parameters may be insufficiently (or not at all) excited, and derive our results for a significantly larger class of nonlinear systems.

\textit{Notation:} The set of integers is denoted by $\mathbb{I}$, the set of all integers greater than or equal to $a$ for any $a \in \mathbb{I}$ by $\mathbb{I}_{\geq a}$, and the set of integers in the interval $[a,b]$ for any $a,b\in \mathbb{I}$ by $\mathbb{I}_{[a,b]}$.
The floor-function applied to some $b\in\mathbb{R}_{\geq0}$ is defined as $\lfloor b \rfloor = \max\{b'\in\mathbb{I}_{\geq0}:b'\leq b \}$.
The $n \times n$ identity matrix is denoted by $I_n$ and the $n \times m$ zero matrix by $0_{n\times m}$, where we omit the indices if the dimension is unambiguous from the context.
The weighted Euclidean norm of a vector $x \in \mathbb{R}^n$ with respect to a positive definite matrix $Q=Q^\top$ is defined as $\|x\|_Q=\sqrt{x^\top Q x}$ with $\|x\|=\|x\|_{I_n}$; the minimal and maximal eigenvalues of $Q$ are denoted by $\underline{\lambda}(Q)$ and $\overline{\lambda}(Q)$, respectively.
For two matrices $A=A^\top$ and $B=B^\top$, we write $A \succeq B$ ($A\succ B$) if $A-B$ is positive semi-definite (positive definite).
For $A,B$ positive definite, the maximum generalized eigenvalue (i.e., the largest scalar $\lambda$ satisfying $\det (A-\lambda B) = 0$) is denoted by $\overline{\lambda}(A,B)$.

\section{Problem setup and preliminaries}
We consider discrete-time systems in the form of
\begin{subequations}
	\label{eq:sys}
	\begin{align}
	x_{t+1} &= f(x_t,u_t,w_t,p),\label{eq:sys_1}\\
	y_t &= h(x_t,u_t,w_t,p),\label{eq:sys_2}
	\end{align}
\end{subequations}
where $t\in\mathbb{I}_{\geq0}$ is the discrete time, $x_t\in\mathbb{R}^n$ is the state at time $t$, $y_t\in\mathbb{R}^p$ is the (noisy) output measurement, $u_t\in\mathbb{R}^m$ is a known input (e.g., the control input), $w_t\in\mathbb{R}^q$ is an unknown generalized disturbance input (representing both process disturbances and measurement noise for the sake of conciseness, which covers the standard setting of independent process disturbance and measurement noise as a special case, cf.~the simulation example in Section~\ref{sec:example}), and $p\in\mathbb{R}^o$ is an unknown but constant parameter.
The nonlinear continuous functions $f:\mathbb{R}^n\times\mathbb{R}^m\times\mathbb{R}^q\times\mathbb{R}^o\rightarrow\mathbb{R}^n$ and $h:\mathbb{R}^n\times\mathbb{R}^m\times\mathbb{R}^q\times\mathbb{R}^o\rightarrow\mathbb{R}^p$ represent the system dynamics and output equation, respectively.
We assume that the unknown true system trajectory satisfies $(x_t,u_t,w_t,p)\in\mathbb{Z}$, $t\in\mathbb{I}_{\geq 0}$,
where
\begin{equation*}	
	\mathbb{Z}:=\{(x,u,w,p)\in\mathbb{X}\times\mathbb{U}\times\mathbb{W}\times\mathbb{P}:f(x,u,w,p)\in\mathbb{X}\}
\end{equation*}
and $\mathbb{X}\subseteq \mathbb{R}^n, \mathbb{U}\subseteq \mathbb{R}^m, \mathbb{W}\subseteq \mathbb{R}^q, \mathbb{P}\subseteq \mathbb{R}^o$ are some known closed sets.
Such constraints typically arise from the physical nature of the system, e.g., non-negativity of partial pressures, mechanically imposed limits, or parameter ranges.
Using this information can significantly improve the estimation results, cf. \cite[Sec.~4.4]{Rawlings2017}.
If no such sets are known \textit{a priori}, they can simply be chosen as $\mathbb{X}= \mathbb{R}^n, \mathbb{U}= \mathbb{R}^m, \mathbb{W}= \mathbb{R}^q, \mathbb{P}= \mathbb{R}^o$.

The overall goal is to compute at each time~$t\in\mathbb{I}_{\geq0}$ the estimates $\hat{x}_t$ and $\hat{p}_t$ of the true values $x_t$ and $p$ using some \textit{a priori} estimates $\hat{x}_0$ and $\hat{p}_0$ and the past measured input-output sequence $\{(u_j,y_j)\}_{j=0}^{t-1}$.
To this end, we require suitable detectability and excitation properties.

\begin{assum}[State detectability]\label{ass:IOSS}
	System~\eqref{eq:sys} admits an i-IOSS Lyapunov function $U:\mathbb{X}\times\mathbb{X}\rightarrow \mathbb{R}_{\geq0}$, that is, there exist matrices $\underline{U},\overline{U},S_{\mathrm{x}},Q_{\mathrm{x}},{R}_{\mathrm{x}}\succ0$ and a constant $\eta_\mathrm{x}\in[0,1)$ such that
	\begin{align}
		&\|x-\tilde{x}\|_{\underline{U}}^2 \leq U(x,\tilde{x}) \leq \|x-\tilde{x}\|_{\overline{U}}^2,\label{eq:IOSS_bounds}\\[1ex]
		&U(f(x,u,w,p),f(\tilde{x},u,\tilde{w},\tilde{p})) \nonumber \\
		&\hspace{3ex} \leq \eta_\mathrm{x} U(x,\tilde{x}) + \|p-\tilde{p}\|_{S_\mathrm{x}}^2 + \|w-\tilde{w}\|_{Q_\mathrm{x}}^2  \nonumber  \\
		& \hspace{3ex} \phantom{\leq} \ + \|h(x,u,w,p)-h(\tilde{x},u,\tilde{w},\tilde{p})\|_{R_\mathrm{x}}^2 \label{eq:IOSS_dissip}
	\end{align}
	for all $(x,u,w,p),(\tilde{x},u,\tilde{w},\tilde{p}) \in \mathbb{Z}$.
\end{assum}

For the stacked input vector $\bar{w}^\top=[w^\top,p^\top]$, Assumption~\ref{ass:IOSS} is equivalent to exponential i-IOSS, which became a standard detectability condition in the context of MHE (for state estimation) in recent years, see, e.g., \cite{Allan2021a,Hu2023,Knuefer2023,Rawlings2017,Schiller2023c}.
This property implies that the difference between any two state trajectories is bounded by the differences of their initial states, their disturbance inputs, their parameters, and their outputs.
Assumption~\ref{ass:IOSS} is not restrictive; in fact, by a straightforward extension of the results from~\cite{Allan2021,Knuefer2023}, it is necessary and sufficient for the existence of robustly stable state estimators if the true parameter is known, and for practically stable state estimators with respect to the parameter error.
Moreover, Assumption~\ref{ass:IOSS} can be verified using LMIs, cf., e.g,~\cite{Schiller2023c}.

\begin{defn}[Set of excited trajectory pairs]\label{def:obs}
	Fix $S_\mathrm{p},P_\mathrm{p},Q_\mathrm{p},R_\mathrm{p}\succ0$ and $\eta_{\mathrm{p}}\in[0,1)$. We define the set containing all excited trajectory pairs of length $T\in\mathbb{I}_{\geq0}$~as
	\begin{align*}
		\mathbb{E}_T := \Big\{&\left(\left\{(x_t,u_t,w_t,p)\right\}_{t=0}^{T-1},\left\{(\tilde{x}_t,u_t,\tilde{w}_t,\tilde{p})\right\}_{t=0}^{T-1}\right)\nonumber \\
		& \in\mathbb{Z}^T\times\mathbb{Z}^T: \nonumber\\
		&x_{t+1}=f(x_t,u_t,w_t,p), \ \tilde{x}_{t+1}=f(\tilde{x}_t,u_t,\tilde{w}_t,\tilde{p}),\nonumber\\
		&y_t=h(x_t,u_t,w_t,p),\ \tilde{y}_t=h(\tilde{x}_t,u_t,\tilde{w}_t,\tilde{p}),\nonumber\\	
		&t\in\mathbb{I}_{[0,T-1]},\nonumber\\
		&\|p-\tilde{p}\|^2_{S_\mathrm{p}} \leq \eta_\mathrm{p}^T\|x_{0}-\tilde{x}_{0}\|^2_{P_\mathrm{p}} 	\nonumber\\
		&\ + \sum_{j={0}}^{T-1}\eta_{\mathrm{p}}^{T-j-1}\Big(\|w_{j}{-\,}\tilde{w}_{j}\|^2_{Q_\mathrm{p}} + \|y_{j}{-\,}\tilde{y}_{j}\|^2_{R_\mathrm{p}}\Big) \Big\}.
	\end{align*}
\end{defn}

The set $\mathbb{E}_T$ contains all trajectory pairs of length $T$ that exhibit a $T$-step distinguishability property with respect to the parameters.
Specifically, for two trajectories that share the same initial state and the same disturbance inputs, if they form a pair contained in the set $\mathbb{E}_T$, it holds that the sum of their output differences is zero if and only if their parameters are the same. In Section~\ref{sec:MHE_PE}, we discuss the relation between detectability of the states (Assumption~\ref{ass:IOSS}), excited trajectory pairs (Definition~\ref{def:obs}), uniform excitation, and a uniform joint detectability condition for both the states and the parameters.

\section{Moving horizon estimation without uniformly persistently excited data}\label{sec:MHE_nonPE}

\subsection{Design}\label{sec:MHE_nonPE_design}
At each time $t\in\mathbb{I}_{\geq 0}$, the proposed MHE scheme considers measured past input-output sequences of the system~\eqref{eq:sys} within a moving horizon of length $N_t = \min\{t, N\}$ for some $N\in\mathbb{I}_{\geq0}$. The current state and parameter estimates are obtained by solving the following nonlinear program:
\begin{subequations}\label{eq:MHE}
	\begin{align}\label{eq:MHE_0}
	&\min_{\hat{x}_{t-N_t|t},\hat{p}_{|t},\hat{w}_{\cdot|t}}\  J_t(\hat{x}_{t-N_t|t},\hat{p}_{|t},\hat{w}_{\cdot|t},\hat{y}_{\cdot|t}) \\ 
	&\text{s.t. }
	\hat{x}_{j+1|t}=f(\hat{x}_{j|t},u_j,\hat{w}_{j|t},\hat{p}_{|t}), \ j\in\mathbb{I}_{[t-N_t,t-1]}, \label{eq:MHE_1} \\	
	&\hspace{0.6cm} \hat{y}_{j|t}=h_{}(\hat{x}_{j|t},u_j,\hat{w}_{j|t},\hat{p}_{|t}),\ j\in\mathbb{I}_{[t-N_t,t-1]}, \label{eq:MHE_2} \\
	&\hspace{0.6cm}(\hat{x}_{j|t},u_j,\hat{w}_{j|t},\hat{p}_{|t})\in\mathbb{Z},\ j\in\mathbb{I}_{[t-N_t,t-1]} \label{eq:MHE_con_x}
	\end{align}
\end{subequations}
for all $t\in\mathbb{I}_{\geq0}$. The decision variables $\hat{x}_{t-N_t|t}$, $\hat{p}_{|t}$, and $\hat{w}_{\cdot|t} = \{\hat{w}_{j|t}\}_{j=t-N_t}^{t-1}$ denote the current estimates of the state at the beginning of the horizon, the parameter, and the disturbance sequence over the horizon, respectively, estimated~at time~$t$.
Given the past input sequence $\{u_j\}_{j=t-N_t}^{t-1}$ applied to system~\eqref{eq:sys}, these decision variables (uniquely) define a sequence of state and output estimates $\{\hat{x}_{j|t}\}_{j=t-N_t}^{t}$ and $\{\hat{y}_{j|t}\}_{j=t-N_t}^{t-1}$ under \eqref{eq:MHE_1} and \eqref{eq:MHE_2}, respectively.
We use the cost function
\begin{align}
	&J_t(\hat{x}_{t-N_t|t},\hat{p}_{|t},\hat{w}_{\cdot|t},\hat{y}_{\cdot|t}) \nonumber \\ 
	&{=\,} \gamma(N_t)\|\hat{x}_{t-N_t|t}{\,-\,}\bar{x}_{t-N_t}\|_{W_{t-N_t}}^2{\hspace{-0.2ex}+\,} \eta_1^{N_t}\|\hat{p}_{|t}{\,-\,}\bar{p}_{t-N_t}\|_{V_{t-N_t}}^2 \nonumber \\
	& \ \ +\sum_{j=1}^{N_t}\eta_2^{j-1}\big(\|\hat{w}_{t-j|t}\|_{Q_{t-j}}^2+\|\hat{y}_{t-j|t}-y_{t-j}\|_{R_{t-j}}^2\big), \label{eq:MHE_objective}
\end{align}
where $\{y_j\}_{j=t-N_t}^{t-1}$ is the measured output sequence of system~\eqref{eq:sys}.
The prior estimates $\bar{x}_{t-N_t}$ and $\bar{p}_{t-N_t}$ as well as the cost function parameters $\gamma(\cdot)$, $\eta_1$, $\eta_2$, $W_t$, $V_t$, $Q_t$, and $R_t$ are defined below.
Note that we consider the prediction form of the estimation problem, i.e., without using the most recent measurement $y_t$ in~\eqref{eq:MHE_objective}, which is commonly done to simplify the notation in the theoretical analysis of MHE schemes, cf.,~e.g.,~\cite{Schiller2023c,Knuefer2023,Allan2021a,Hu2023}, and see also~\cite[Ch.~4]{Rawlings2017} for a discussion on this topic.
We impose the following assumption on the cost function weights.
\begin{assum}[Cost function]\label{ass:bounds2}
	The discount parameters $\gamma$, $\eta_1$, $\eta_2$ satisfy
	\begin{align}
		&\gamma(s) = \eta_{\mathrm{x}}^s +\overline{\lambda}(P_\mathrm{p},\overline{U})\eta_{\mathrm{p}}^s,\  s\geq0, \label{eq:gamma}\\
		&\eta_1 \in (\max\{\eta_{\mathrm{x}},\eta_{\mathrm{p}}\},1), \\
		&\eta_2 \in [\max\{\eta_{\mathrm{x}},\eta_{\mathrm{p}}\},1).\label{eq:eta_2}
	\end{align}
	There exist matrices $\overline{W},\underline{V},\overline{V},\overline{Q},\overline{R}\succ0$ such that
	\begin{align}
		&2\overline{U} \preceq W_t \preceq \overline{W},\label{eq:bound_W2}\\
		&2S_\mathrm{p}\preceq 2\underline{V} \preceq  V_t \preceq \overline{V},\label{eq:bound_V2}\\
		&2(Q_{\mathrm{x}}+Q_\mathrm{p}) \preceq  Q_t \preceq \overline{Q},\label{eq:bound_Q2}\\
		&R_{\mathrm{x}}+R_\mathrm{p} \preceq  R_t \preceq \overline{R} \label{eq:bound_R2}
	\end{align}
	uniformly for all $t\in\mathbb{I}_{\geq0}$.
\end{assum}
Assumption~\ref{ass:bounds2} ensures large tuning capabilities of the cost function~\eqref{eq:MHE_objective} while satisfying certain relations to the detectability and excitation properties from Assumption~\ref{ass:IOSS} and Definition~\ref{def:obs}. This is conceptually similar to the recent MHE literature (for state estimation) and typically permits a less conservative stability analysis due to the structural similarities between the detectability condition, MHE, and stability, cf., e.g., \cite[Sec.~III.D]{Schiller2023c} and compare also \cite{Hu2023,Knuefer2023}.
Potential time dependency of the weighting matrices in~\eqref{eq:bound_W2}--\eqref{eq:bound_R2} can be used to incorporate additional knowledge (e.g., by choosing Kalman filter covariance update laws \cite{Qu2009,Rao2003}), which can be beneficial in practice to increase estimation performance.
Assumption~\ref{ass:bounds2} implies that the cost function~\eqref{eq:MHE_objective} is radially unbounded in the decision variables, which together with continuity of $f$ and $h$ ensures that the estimation problem~\eqref{eq:MHE} and \eqref{eq:MHE_objective} admits a globally optimal solution at any time $t\in\mathbb{I}_{\geq0}$, cf.~\cite[Sec.~4.2]{Rawlings2017}.
We denote a minimizer by $(\hat{x}^*_{t-N_t|t},\hat{p}^*_{|t},\hat{w}_{\cdot|t}^*)$, and the corresponding optimal state sequence by $\{\hat{x}^*_{j|t}\}_{j=t-N_t}^t$.
The resulting state and parameter estimates at time $t\in\mathbb{I}_{\geq0}$ are then given by $\hat{x}_t = \hat{x}^*_{t|t}$ and $\hat{p}_t =  \hat{p}^*_{|t}$, yielding the estimation error
\begin{equation}\label{eq:MHE_error}
	e_t^\top
	= \big[e_{\mathrm{x},t}^\top,\ e_{\mathrm{p},t}^\top\big]
	= \big[(\hat{x}_t-x_t)^\top,\ (\hat{p}_t-p)^\top\big].
\end{equation}
It remains to define suitable update laws for the prior estimates to ensure a proper regularization of the cost function~\eqref{eq:MHE_objective}. For the state prior, we select $\bar{x}_{t}=\hat{x}_{t}$ (which is typically called the \textit{filtering prior}, cf.~\cite[Sec.~4.3]{Rawlings2017}).
For the parameter prior, we propose the following update law:	
\begin{equation}\label{eq:pbar}
	\bar{p}_{t} =
	\begin{cases}
		\hat{p}_t,& \text{if $t\in\mathbb{I}_{\geq N}$ and $X_{t}\in\mathbb{E}_{N}$},\\
		\bar{p}_{t-N_t}, & \text{otherwise}
	\end{cases}
\end{equation}
with $\bar{p}_0 = \hat{p}_0$, where $X_{t}:=\Big(\{(\hat{x}_{j|t}^*,u_j,\hat{w}_{j|t}^*,\hat{p}_{|t}^*)\}_{j=t-N_t}^{t-1},\allowbreak \{({x}_{j},u_j,{w}_{j},p)\}_{j=t-N_t}^{t-1}\Big)$ is the pair of the (unknown) true and the currently optimal trajectory.
The update in~\eqref{eq:pbar} depends on the currently present level of excitation; in Section~\ref{sec:dIOSS}, we propose a suitable method to practically check whether $X_{t}\in\mathbb{E}_{N}$ online.
In the following, we show how the horizon length $N$ must be chosen so that the estimation error~\eqref{eq:MHE_error} exhibits a certain robust stability property that is valid regardless of the parameter excitation.

\subsection{Stability Analysis}\label{sec:MHE_nonPE_stability}
We consider the two Lyapunov function candidates:
\begin{align}
	&\Gamma_1(t,\hat{x},x,\hat{p},p) = U(\hat{x},x) + \|\hat{p}-p\|_{V_t}^2, \label{eq:candidate}\\
	&\Gamma_2(\hat{x},x,\hat{p},p) = U(\hat{x},x) + c\|\hat{p}-p\|_{\overline{V}}^2, \ c\geq 1, \label{eq:candidate2}
\end{align}
where $U$ is from Assumption~\ref{ass:IOSS} and $V_t\preceq\overline{V}$ is from~\eqref{eq:MHE_objective} under Assumption~\ref{ass:bounds2}.
The following two auxiliary results establish fundamental properties of $\Gamma_1$ and $\Gamma_2$ for the two cases where the current level of excitation is too low ($X_t \notin E_N$, Lemma~\ref{lem:nonPE}) or sufficiently high ($X_t \in \mathbb{E}_N$, Lemma~\ref{lem:PE}); the proofs can be found in Appendix~\ref{sec:appendix}.

\begin{lem}\label{lem:nonPE}
	Let Assumption~\ref{ass:IOSS} hold. Consider the MHE scheme~\eqref{eq:MHE} with the cost function \eqref{eq:MHE_objective} satisfying Assumption~\ref{ass:bounds2}.
	Assume that $t\in\mathbb{I}_{[0,N-1]}$ or $t\in\mathbb{I}_{\geq N}$ and $X_t\notin\mathbb{E}_N$. Then, it holds that
	\begin{align}
		&\Gamma_1(t,\hat{x}_t,x_t,\hat{p}_t,p)\nonumber\\
		&\leq \eta_{\mathrm{1}}^{-N}c_{1}(N_t)(\eta_\mathrm{x}^{N_t}+\gamma(N_t))\|\bar{x}_{t-N_t}-x_{t-N_t}\|_{\overline{W}}^2\nonumber\\
		& \quad + 2c_{1}(N_t)\eta_1^{-N}\eta_1^{N_t}\|\bar{p}_{t-N_t}-p\|^2_{\overline{V}}\nonumber\\
		& \quad + 2c_{1}(N_t)\eta_{\mathrm{1}}^{-N}\sum_{j={1}}^{N_t}\eta_2^{j-1}\|w_{t-j}\|_{\overline{Q}}^2,\label{eq:proof_case_1}
	\end{align}
	where
	\begin{equation}
		c_1(s):=\bar{\lambda}(S_\mathrm{x},\underline{V})\frac{1-\eta_\mathrm{x}^{s}}{1-\eta_\mathrm{x}} + \overline{\lambda}(\overline{V},\underline{V}),\ s\geq 0. \label{eq:c1_s}
	\end{equation}
\end{lem}

\begin{lem}\label{lem:PE}
	Let Assumption~\ref{ass:IOSS} hold. Consider the MHE scheme~\eqref{eq:MHE} with the cost function \eqref{eq:MHE_objective} satisfying Assumption~\ref{ass:bounds2}.
	Assume that $X_t\in\mathbb{E}_N$ for some $t\in\mathbb{I}_{\geq N}$. Then, it holds that
	\begin{align}	
		\Gamma_2(\hat{x}_t,x_t,\hat{p}_t,p) \leq&\ 
		\mu^N\Gamma_1(t-N,\bar{x}_{t-N},x_{t-N},\bar{p}_{t-N},p)\nonumber\\
		&\ + 2c_{2}(c,N)\sum_{j={1}}^{N}{\eta}_2^{j-1}\|w_{t-j}\|_{\overline{Q}}^2,\label{eq:lem_PE2}
	\end{align}
	for all $c\geq 1$, where
	\begin{equation}\label{eq:mu}
		\mu := \max\left\{\sqrt[N]{2\overline{\lambda}(\overline{W},\underline{U})c_2(c,N)\gamma(N)},\sqrt[N]{c_2(c,N)}\eta_1\right\}
	\end{equation}
	and, for any $c\geq 1$ and $s\geq0$,
	\begin{equation}\label{eq:c}
		c_2(c,s) := c\overline{\lambda}(\overline{V},S_{\mathrm{p}}) + 	\overline{\lambda}(S_\mathrm{x},S_{\mathrm{p}})({1-\eta_{\mathrm{x}}^s})/(1-\eta_{\mathrm{x}}).
	\end{equation}
\end{lem}

Now, let
	\begin{equation}\label{eq:rho}
		\rho := \max\left\{\eta_1^{-N}c_{1}(N)(\eta_\mathrm{x}^N{\,+\,}\gamma(N))\overline{\lambda}(\overline{W},\underline{U}),\eta_2^N\right\}
	\end{equation}	
and $c$ be such that
	\begin{equation}
		c =  {2}c_{1}(N)/(1-\rho) + 1 \label{eq:c_def} 
	\end{equation}
with $c_{1}$ from~\eqref{eq:c1_s}.
The robustness guarantees for the proposed MHE scheme require satisfaction of the following conditions on the horizon length $N$:
\begin{align}
	&2\overline{\lambda}(\overline{W},\underline{U})c_2(c,N)\gamma(N) <1,\label{eq:contraction_2}\\
	&c_2(c,N)\eta_1^N<1,\label{eq:contraction_3}\\
	&\eta_1^{-N}c_{1}(N)(\eta_\mathrm{x}^N+\gamma(N))\overline{\lambda}(\overline{W},\underline{U}) < 1,\label{eq:contraction_1}
\end{align}
where $c_1(s)$ and $\gamma(s)$ are from~\eqref{eq:c1_s} and~\eqref{eq:gamma}, respectively.
The conditions~\eqref{eq:contraction_2} and~\eqref{eq:contraction_3} imply that $\mu\in[0,1)$ in~\eqref{eq:mu} and hence ensure contraction of the state and parameter estimation error in case the excitation condition used in~\eqref{eq:pbar}, i.e., $X_t\in\mathbb{E}_N$, is met.
Condition~\eqref{eq:contraction_1} implies $\rho\in[0,1)$ in~\eqref{eq:rho} and $c\geq1$ in~\eqref{eq:c_def} and hence ensures boundedness of the estimation error in case the excitation condition is not met.
Under Assumption~\ref{ass:bounds2}, there always exists $N$ sufficiently large such that the contraction conditions~\eqref{eq:contraction_2}--\eqref{eq:contraction_1} are satisfied (to see this, note that the left-hand side of each of these conditions can be bounded by a function that exponentially decays to zero as $N\rightarrow\infty$, which follows by invoking \eqref{eq:gamma}--\eqref{eq:eta_2} and uniform boundedness of $c_1$ and $c_2$ from \eqref{eq:c1_s} and~\eqref{eq:c}).

At each time $t\in\mathbb{I}_{\geq0}$, we split the interval $[0,t]$ into sub-intervals of length $N$ and the remainder $l=t-\lfloor{t/N}\rfloor N$:
\begin{equation}\label{eq:times}
	t = l + \sum_{m=1}^{k} (i_m+1)N + jN,
\end{equation}
where $k\in\mathbb{I}_{\geq0}$, $i_m\in\mathbb{I}_{[1,k]}$ for $k\in\mathbb{I}_{\geq1}$, and $j\in\mathbb{I}_{\geq0}$ are defined as follows.
To this end, let the set $\mathcal{T}_t := \left\{\tau\in\mathbb{I}_{[N,t]} : t-\tau-N\left\lfloor\frac{t-\tau}{N}\right\rfloor=0, X_{\tau}\in\mathbb{E}_{N}\right\}$ contain all past time instances from the set $\{t, t-N, t-2N, ... \}$ at which the excitation condition used in~\eqref{eq:pbar} was met (in the following also referred to as \emph{PE horizons} for simplicity).
The variable $k$ denotes the total number of PE horizons that occurred until the current time $t$ and is defined as the cardinality of $\mathcal{T}_t$, i.e., $k:= |\mathcal{T}_t|$.
Suppose that $k\in\mathbb{I}_{\geq1}$. The sequence $\{t_m\}_{m=1}^k$ contains time instances corresponding to PE horizons, where $t_1 := \max\{\tau\in\mathcal{T}_t\}$ and $t_{m+1} = \max\{\tau\in\mathcal{T}_t:\tau<t_m\}$ for $m\in\mathbb{I}_{[1,k-1]}$ if $k\in\mathbb{I}_{\geq 2}$.
The sequence $\{i_m\}_{m=1}^k$ denotes the numbers of non-PE horizons (i.e., past time instances where the excitation condition~\eqref{eq:pbar} was not met) between two successive times $t_m$ and $t_{m+1}$ with $i_{m} = ({t_m-N-t_{m+1}})/{N}$ for $m\in\mathbb{I}_{[1,k-1]}$ if $k\in\mathbb{I}_{\geq 2}$, and\footnote{
	Note that $\{t_m\}_{m=1}^k, \{i_m\}_{m=1}^k$ do not have to be formally defined for $k=0$, as this yields an empty sum in \eqref{eq:times}.
} $i_k = (t_k-N-l)/N$.
Finally, $j$ stands for the number of non-PE horizons that occurred between time $t$ and $t_1$ if $k\geq1$ (and between time $t$ and $l$ if $k=0$), i.e., $j := ({t-\max\{\tau\in\mathcal{T}_t,l\}})/{N}$.
Overall, the partitioning in~\eqref{eq:times} allows us to theoretically cover occasional occurrence of PE horizons, which of course includes the special cases in which PE horizons never occur ($k=0$), or in which all horizons are PE ($j=0$, $i_m=0$ for all $m=1,...,k$).

We are now in a position to state our main result. Our key argument is that $\Gamma_2$~\eqref{eq:candidate2} decreases from $t_m$ to $t_{m+1}$.

\begin{thm}\label{thm:nonPE}
	Let Assumption~\ref{ass:IOSS} hold. Consider the MHE scheme~\eqref{eq:MHE} with the cost function \eqref{eq:MHE_objective} satisfying Assumption~\ref{ass:bounds2}.	
	Suppose that the horizon length $N$ satisfies \eqref{eq:contraction_2}--\eqref{eq:contraction_1}.
	Then, it holds that
	\begin{align}
		&\frac{1}{C_0}\Gamma_1(t,\hat{x}_{t},x_{t},\hat{p}_{t},p)\label{eq:thm_nonPE}\\
		&\leq \mu^{kN}\Big(C_1\tilde{\eta}^l\|\hat{x}_{0}-x_{0}\|_{\overline{W}}^2 + C_2\eta_1^l \|\hat{p}_{0}-p\|^2_{\overline{V}}\Big)\nonumber\\
		&\hspace{2ex} + \mu^{kN}\sum_{r={1}}^{l}\eta_2^{r-1}\|w_{l-r}\|_{{Q}}^2  +  \sum_{r={1}}^{jN}\rho_N^{r-1}\|w_{t-r}\|_{{Q}}^2 \nonumber\\
		&\hspace{2ex} +\sum_{m=1}^k \hspace{-0.5ex}\mu^{(m-1)N}\hspace{-1ex}\sum_{r={1}}^{(i_m+1)N} \hspace{-1.5ex}\bar{\mu}^{r-1}\|w_{t-jN-\sum_{q=1}^{m-1}(i_q+1)N-r}\|_{Q}^2\nonumber
	\end{align}
	for all $t\in\mathbb{I}_{\geq0}$ and all $\hat{x}_0,x_0\in\mathbb{X}$, all $\hat{p}_0,p\in\mathbb{P}$, and every disturbance sequence $\{w_r\}_{r=0}^{\infty}\in\mathbb{W}^\infty$, where	$\tilde{\eta} = \max\{\eta_\mathrm{x},\eta_\mathrm{p}\}$, $\bar{\mu}=\max\{\mu,\rho_N\}$, $\rho_N=\sqrt[N]{\rho}$ with $\rho$ from~\eqref{eq:rho}, and
	\begin{align}
		C_0 &:=c\overline{\lambda}(\overline{V},\underline{V}),\label{eq:C0_def}\\
		C_1 &:= \eta_{\mathrm{1}}^{-N}c_{1}(N)(2+ \overline{\lambda}(P_\mathrm{p},\overline{U})),\label{eq:C1_def}\\
		C_2 &:= (2c_1(N)+\overline{\lambda}(\overline{V},\underline{V})^{-1})\eta_1^{-N}, \label{eq:C2_def}\\
		Q&:=\max\{\eta_{\mathrm{1}}^{-N}c_{1}(N),c_2(c,N)\}2\overline{Q}.\label{eq:Q_def}
	\end{align}
\end{thm}

Before proving Theorem~\ref{thm:nonPE}, we want to highlight some key properties.
\begin{enumerate}
	\item Theorem~\ref{thm:nonPE} provides a bound on the state and parameter estimation error\footnote{
		By the definition of $\Gamma_1$ in~\eqref{eq:candidate} and the lower bound in~\eqref{eq:IOSS_bounds}.
	} that is valid regardless of the parameter excitation; it also applies if the excitation condition in~\eqref{eq:pbar} is never met (which corresponds $k=0$ in~\eqref{eq:thm_nonPE}) and constitutes a bounded-disturbance bounded-estimation-error property.
	
	\item Satisfaction of the excitation condition in~\eqref{eq:pbar} for some $t\in\mathbb{I}_{\geq N}$ (leading to non-zero values of $k$ in~\eqref{eq:thm_nonPE}) always improve the error bound~\eqref{eq:thm_nonPE} with respect to the initial estimation error.
	
	\item If $k\rightarrow\infty$ for $t\rightarrow\infty$, then the estimation error converges to a ball centered at the origin with the radius defined by the true disturbances.
\end{enumerate}

\begin{pf*}{Proof of Theorem~\ref{thm:nonPE}.}
	We prove the statement in five parts---namely, by deriving bounds 1) on $\Gamma_1(t,\cdot)$ involving data from $[t-jN,t-1]$; 2) on $\Gamma_2(\cdot)$ involving data from $[t_2,t_1-1]$; 3) on $\Gamma_2(\cdot)$ involving data from $[t_k,t_1-1]$; 4) on $\Gamma_2(\cdot)$ involving data from $[0,l-1]$; 5) on $\Gamma_1(t,\cdot)$ involving all available data from $[0,t-1]$.
	
	\textbf{Part 1.} We define $\tilde{\Gamma}_1(t) := \Gamma_1(t,\hat{x}_{t},x_{t},\hat{p}_{t},p)$ for notational brevity and assume that $j\in\mathbb{I}_{\geq1}$.
	By applying Lemma~\ref{lem:nonPE} together with the fact that $\|\bar{x}_{t-N}-x_{t-N}\|_{\overline{W}}^2\leq \overline{\lambda}(\overline{W},\underline{U})U(\bar{x}_{t-N},x_{t-N})$ by \eqref{eq:IOSS_bounds} and the contraction condition~\eqref{eq:contraction_1}, we obtain
	\begin{align}
		\tilde{\Gamma}_1(t)\leq &\
		\rho U(\bar{x}_{t-N},x_{t-N}) + c_1' \|\bar{p}_{t-N}-p\|^2_{\overline{V}}  \nonumber\\
		&\ + \sum_{r={1}}^{N}{\eta}_2^{r-1}\|w_{t-r}\|_{Q}^2, \label{eq:proof_1}
	\end{align}
	where $c_1'=2c_{1}(N)$, $2c_{1}(N)\eta_{\mathrm{1}}^{-N}\overline{Q}\preceq Q$ with $Q$ from~\eqref{eq:Q_def}, and $\rho$ satisfies $\rho<1$.
	In~\eqref{eq:proof_1}, note that $\bar{x}_{t-N}=\hat{x}_{t-N}$ and $\bar{p}_{t-N} = \bar{p}_{t-qN} = \bar{p}_{t-jN}$ for all $q = 1,...,j$ due to the update rule~\eqref{eq:pbar}.
	Since $U(\hat{x}_{t-N},x_{t-N})\leq\Gamma_1(t-N,\hat{x}_{t-N},x_{t-N},\hat{p}_{t-N},p)$ 	by~\eqref{eq:candidate}, we can recursively apply \eqref{eq:proof_1} for $j$ times, yielding
	\begin{align}
		\tilde{\Gamma}_1(t) \leq&\ \rho^jU(\hat{x}_{t-jN},x_{t-jN}) + \sum_{q={1}}^{j} \rho^{q-1}c_1' \|\bar{p}_{t-jN}-p\|^2_{\overline{V}}\nonumber  \\
		&\ +  \sum_{q={1}}^{j} \rho^{q-1} \sum_{r={1}}^{N}\eta_2^{r-1}\|w_{t-(q-1)N-r}\|_{{Q}}^2.\label{eq:proof_P1}
	\end{align}
	Using $\rho_N:=\sqrt[N]{\rho}\geq\eta_2$ (the latter inequality follows by the definition of $\rho$ from~\eqref{eq:rho}), the geometric series, and $c_1'/(1-\rho)=2c_1(N)/(1-\rho)<c$ by~\eqref{eq:c_def}, we have that
	\begin{align}
		\tilde{\Gamma}_1(t)\leq&\ \Gamma_2(\hat{x}_{t-jN},x_{t-jN},\bar{p}_{t-jN},p) +  \sum_{r={1}}^{jN}\rho_N^{r-1}\|w_{t-r}\|_{{Q}}^2.\label{eq:proof_P1_res}
	\end{align}
	\textbf{Part 2.} Assume that $k\in\mathbb{I}_{\geq1}$. Then, $t-jN=t_1$ corresponds to the most recent PE horizon where $X_{t_1}\in\mathbb{E}_N$ and $\bar{p}_{t-jN} = \hat{p}_{t_1}$ by \eqref{eq:pbar}. Invoking Lemma~\ref{lem:PE} yields
	\begin{align}
		&\Gamma_2(\hat{x}_{t-jN},x_{t-jN},\bar{p}_{t-jN},p) = \Gamma_2(\hat{x}_{t_1},x_{t_1},\hat{p}_{t_1},p)\nonumber\\
		&\leq \mu^N\Gamma_1(t_1-N,\bar{x}_{t_1-N},{x}_{t_1-N},\bar{p}_{t_1-N},p)\nonumber\\
		&\quad + \sum_{r={1}}^{N}{\eta}_2^{r-1}\|w_{t_1 - r}\|_{Q}^2, \label{eq:proof_P2_1}
	\end{align}
	where we have used that $2c_{2}(c,N)\overline{Q}\preceq Q$ with $Q$ from~\eqref{eq:Q_def}.
	By the definition of $\Gamma_1$ from \eqref{eq:candidate}, we obtain
	\begin{align}
		&\Gamma_1(t_1-N,\bar{x}_{t_1-N},{x}_{t_1-N},\bar{p}_{t_1-N},p) \nonumber \\
		&\leq  \Gamma_1(t_1-N,\hat{x}_{t_1-N},{x}_{t_1-N},\hat{p}_{t_1-N},p)  \nonumber\\
		&\quad + \|\bar{p}_{t_1-(i_1+1)N}-p\|_{\overline{V}}, \label{eq:proof_P2_2}
	\end{align}
	where we have used that $\bar{x}_{t_1-N}=\hat{x}_{t_1-N}$ and $\bar{p}_{t_1-N} = \bar{p}_{t_1-qN} = \bar{p}_{t_1-(i_1+1)N}$ for all $q = 1,...,i_1+1$.
	In the following, consider $k\in\mathbb{I}_{\geq2}$.
	Then, $\bar{p}_{t_1-(i_1+1)N}=\hat{p}_{t_2}$.
	Using a similar argument as in~\eqref{eq:proof_P1}, the geometric series, and the definition of $\rho_N$, we have that
	\begin{align}
		&\Gamma_1(t_1-N,\hat{x}_{t_1-N},x_{t_1-N},\hat{p}_{t_1-N},p)\nonumber \\
		&\leq \rho^{i_1}U(\hat{x}_{t_2},x_{t_2}) + \frac{c_1'}{1-\rho} \|\hat{p}_{t_2}-p\|^2_{\overline{V}}\nonumber  \\
		&\quad + \sum_{r={1}}^{i_1N}\rho_N^{r-1}\|w_{t_1-N-r}\|_{{Q}}^2. \label{eq:proof_P2_3}
	\end{align}
	Combining \eqref{eq:proof_P2_1}--\eqref{eq:proof_P2_3} with the fact that $c=c_1'/(1-\rho)+1$ and the definition of $\bar{\mu}:=\max\{\mu,\rho_N\}$, we obtain
	\begin{align}
		&\Gamma_2(\hat{x}_{t_1},x_{t_1},\hat{p}_{t_1},p)\nonumber\\
		&\leq \mu^N\Gamma_2(\hat{x}_{t_2},x_{t_2},\hat{p}_{t_2},p) + \sum_{r={1}}^{(i_1+1)N}\bar{\mu}^{r-1}\|w_{t_1-r}\|_{{Q}}^2. \label{eq:proof_P2_res}
	\end{align}
	\textbf{Part 3.} Suppose that $k\in\mathbb{I}_{\geq2}$.
	By applying \eqref{eq:proof_P2_res} recursively for all $m\in\mathbb{I}_{[1,k-1]}$ and the fact that $t-jN - \sum_{m=1}^{k}(i_m+1)N = l$ by \eqref{eq:times}, we can infer that
	\begin{align}
		&\Gamma_2(\hat{x}_{t-jN},x_{t-jN},\bar{p}_{t-jN},p)\nonumber\\ 
		&\leq \mu^{kN} \Gamma_2(\hat{x}_{l},x_{l},\bar{p}_{l},p)\label{eq:proof_P3_res}\\
		&+\sum_{m=1}^k \mu^{(m-1)N} \hspace{-1ex}\sum_{r={1}}^{(i_m+1)N}\hspace{-1ex} \bar{\mu}^{r-1}\|w_{t-jN-\sum_{q=1}^{m-1}(i_q+1)N-r}\|_{Q}^2. \nonumber
	\end{align}
	\textbf{Part 4.} By the definition of $\Gamma_2$ from~\eqref{eq:candidate2}, it follows that
	\begin{align}
		&\Gamma_2(\hat{x}_l,x_l,\bar{p}_l,p) = U(\hat{x}_l,x_l) + c\|\bar{p}_l-p\|_{\overline{V}} + c\|\hat{p}_l-p\|_{\overline{V}}\nonumber\\
		&\hspace{2ex} \leq C_0(\Gamma_1(l,\hat{x}_l,x_l,\hat{p}_l,p) + \overline{\lambda}(\overline{V},\underline{V})^{-1}\|\bar{p}_l-p\|_{\overline{V}}) \label{eq:proof_P4_1}
	\end{align}
	with $C_0$ from~\eqref{eq:C0_def}, and where $\bar{p}_l = \hat{p}_0$ by \eqref{eq:pbar}.
	By using Lemma~\ref{lem:nonPE} with $\bar{x}_0=\hat{x}_0$ and $\bar{p}_0=\hat{p}_0$, we obtain
	\begin{align}
		&\Gamma_1(l,\hat{x}_l,x_l,\hat{p}_l,p)\nonumber\\
		&\leq c_{1}(N)\eta_{\mathrm{1}}^{-N}(\eta_\mathrm{x}^{l}+\gamma(l))\|\hat{x}_{0}-x_{0}\|_{\overline{W}}^2\nonumber\\
		& \quad + c_1'\eta_{\mathrm{1}}^{-N}\eta_1^l\|\hat{p}_{0}-p\|^2_{\overline{V}}+ \sum_{r={1}}^{l}\eta_2^{r-1}\|w_{l-r}\|_{Q}^2, \label{eq:proof_P4_2}
	\end{align}
	where we have used the definitions of $c'$ and $Q$ together with the facts that $l<N$ and $c_1(s)$ is monotonically increasing in $s$.
	From \eqref{eq:proof_P4_1} and \eqref{eq:proof_P4_2} and the definitions of $C_1$ and $C_2$ from~\eqref{eq:C1_def} and~\eqref{eq:C2_def}, we can infer that
	\begin{align}
		\Gamma_2(\hat{x}_{l},x_{l},\bar{p}_{l},p) {\,\leq}&\, C_0\Big(C_1\tilde{\eta}^l\|\hat{x}_{0}{-}x_{0}\|_{\overline{W}}^2 {\,+\,} C_2\eta_1^l \|\hat{p}_{0}{-\,}p\|^2_{\overline{V}}\nonumber \\
		&\ + \sum_{r={1}}^{l}\eta_2^{r-1}\|w_{l-r}\|_{{Q}}^2\Big). \label{eq:proof_P4_res}
	\end{align}
	\textbf{Part 5.} The property in~\eqref{eq:thm_nonPE} follows by combining \eqref{eq:proof_P1_res}, \eqref{eq:proof_P3_res}, and \eqref{eq:proof_P4_res}, using that $C_0>1$, and noting that the result holds for all $l\in\mathbb{I}_{[0,N-1]}$, $k\in\mathbb{I}_{\geq0}$, and $j\in\mathbb{I}_{\geq0}$ (i.e., for all $t\in\mathbb{I}_{\geq0}$), which finishes this proof. \qed
\end{pf*}

\begin{rem}\label{rem:est_p}
	The update rule~\eqref{eq:pbar} leads to a certain periodic behavior of the estimation error and its theoretical bounds.
	In particular, while accurate estimates and error bounds propagate over an integer multiple of $N$ time steps, this has no effect on the estimates and bounds in between.
	A practical solution to avoid propagation of poor parameter estimates is to just select $\hat{p}_t$ as the most recent estimate that was computed using PE data, i.e., $\hat{p}_t = \hat{p}^*_{|\tau}, \tau=\max\{\tau\in\mathbb{I}_{[0,t]}, X_\tau\in\mathbb{E}_{N_t}\}$.
	In this case, the above developed theoretical guarantees are still valid: the error $e_{\mathrm{p},t}=\hat{p}_t-p$ is bounded by~\eqref{eq:thm_nonPE} at time $\tau$ due to the fact that $\underline{\lambda}(\underline{V})\|e_{\mathrm{p},\tau}\|^2\leq \Gamma_1(\tau,\hat{x}_\tau,x_\tau,\hat{p}_\tau,p)$.
\end{rem}

If the time between two consecutive PE horizons can be uniformly bounded for all times, Theorem~\ref{thm:nonPE} specializes to robust global exponential stability of the estimation error.

\begin{cor}\label{cor:kappa}
	Let the conditions of Theorem~\ref{thm:nonPE} be satisfied. Assume that there exists a constant $\kappa\geq0$ such that $j\leq \kappa$ and $i_m\leq \kappa$ for all $m\in\mathbb{I}_{[1,k]}$ if $k\in\mathbb{I}_{\geq1}$ uniformly for all $t\in\mathbb{I}_{\geq0}$. Then, the joint estimation error~\eqref{eq:MHE_error} is uniformly robustly globally exponentially stable, that is, there exist $K_1,K_2\geq0$ and $\lambda_1,\lambda_2\in[0,1)$ such that
	\begin{align}\label{eq:RGES}
		\|e_t\| \leq \max\left\{K_1\lambda_1^t \|e_0\|, K_2 \max_{r\in[1,t]} \lambda_2^{r-1}\|w_{t-r}\|\right\}
	\end{align}	
	for all $t\in\mathbb{I}_{\geq0}$ and all $\hat{x}_0,x_0\in\mathbb{X}$, all $\hat{p}_0,p\in\mathbb{P}$, and every disturbance sequence $\{w_r\}_{r=0}^{\infty}\in\mathbb{W}^\infty$.
\end{cor}

\begin{pf}
	Consider~\eqref{eq:thm_nonPE}. Define $\mu_\kappa:=\bar{\mu}^{\frac{1}{\kappa+1}}$ with $\bar{\mu}\geq \max\{\mu, \rho_N\}$ from Theorem~\eqref{thm:nonPE}.
	Since $j$ and $i_m$ are uniformly bounded by $\kappa$ for all $m\in\mathbb{I}_{[1,k]}$ and $k\in\mathbb{I}_{\geq1}$, we can write that
	\begin{equation}
		\mu^{sN} \leq \mu_\kappa^{s(\kappa+1)N} \leq \mu_\kappa^{\sum_{m=1}^s(i_m+1)N} \label{eq:proof_kappa_0}
	\end{equation}
	for all $s \in\mathbb{I}_{[0,k]}$.
	Since, $1\leq\mu_\kappa^{-\kappa N} \mu_\kappa^{qN}$ for all $q\in\{j,\{i_m\}_{m=1}^k\}$ and $\mu_\kappa\geq\tilde\eta$, we can also infer that
	\begin{equation}
		\mu^{kN}\eta_1^l\leq\mu^{kN}\tilde{\eta}^l\leq\mu_\kappa^{-\kappa N}\mu_\kappa^{jN}\mu^{kN}\mu_\kappa^{l} \stackrel{\eqref{eq:proof_kappa_0}}{\leq} \mu_\kappa^{-\kappa N}\mu_\kappa^t. \label{eq:proof_kappa_1}
	\end{equation}
	In addition, we can write that
	\begin{align}
		&\sum_{m=1}^k \hspace{-0.5ex}\mu^{(m-1)N}\hspace{-1ex}\sum_{r={1}}^{(i_m+1)N} \hspace{-1.5ex}\bar{\mu}^{r-1}\|w_{t-jN-\sum_{q=1}^{m-1}(i_q+1)N-r}\|_{Q}^2 \nonumber\\
		&\leq \sum_{r={1}}^{t-jN-l} \hspace{-1.5ex}\bar{\mu}^{r-1}\|w_{t-jN-r}\|_{Q}^2\label{eq:proof_kappa_2}
	\end{align}
	by~\eqref{eq:proof_kappa_0} with $s=m-1$ and~\eqref{eq:times}.
	Hence, from Theorem~\ref{thm:nonPE}, the definition of $\mu_\kappa$, and~\eqref{eq:proof_kappa_0}--\eqref{eq:proof_kappa_2}, we obtain
	\begin{align*}
		&\frac{\mu_\kappa^{\kappa N}}{C_0}\Gamma_1(t,\hat{x}_t,x_t,\hat{p}_t,p)\leq C_1\mu_\kappa^{t}\|\hat{x}_{0}-x_{0}\|_{\overline{W}}^2\\
		&\hspace{16ex} + C_2\mu_\kappa^{t}\|\hat{p}_{0}-p\|^2_{\overline{V}} +\sum_{r={1}}^{t}\mu_\kappa^{r-1}\|w_{t-r}\|_{{Q}}^2.
	\end{align*}
	Using the definition of $\Gamma_1$ from \eqref{eq:candidate}, the estimation error \eqref{eq:MHE_error} satisfies
	\begin{equation}
		\|e_t\|^2 \leq \tilde{K}_1\mu_\kappa^{t} \|e_0\|^2\label{eq:proof_kappa_3} + \tilde{K}_2\sum_{r={1}}^{t}\mu_\kappa^{r-1}\|w_{t-r}\|^2,
	\end{equation}
	with $\tilde{K}_1 := \tilde{K}_3 \max\{C_1\overline{\lambda}(\overline{W}),C_2\overline{\lambda}(\overline{V})\}$, $\tilde{K}_2 :=  \tilde{K}_3 \overline{\lambda}(Q)$, and $\tilde{K}_3:=C_0(\mu_\kappa^{\kappa N}\min\{\underline{\lambda}(\underline{U}),\underline{\lambda}(\underline{V})\})^{-1}$.
	By the geometric series, we have that
	\begin{equation}
		\sum_{r={1}}^{t}\mu_\kappa^{r-1}\|w_{t-r}\|^2 \leq \frac{1}{1-\sqrt{\mu_\kappa}} \max_{r\in[1,t]} \sqrt{\mu_\kappa}^{r-1}\|w_{t-r}\|^2.\label{eq:proof_kappa_4}
	\end{equation}	
	By taking the square root of~\eqref{eq:proof_kappa_3} and using \eqref{eq:proof_kappa_4}, we obtain~\eqref{eq:RGES} with $K_1 = \sqrt{2\tilde{K}_1}$, $K_2 =  \sqrt{2\tilde{K}_2/(1-\sqrt{\mu_\kappa})}$, $\lambda_1 = \sqrt{\mu_\kappa}$, and $\lambda_2=\sqrt[4]{\mu_\kappa}$, which finishes this proof. \qed
\end{pf}

\subsection{Special case: uniform persistent excitation}\label{sec:MHE_PE}

In the following, we consider the special case in which the excitation condition in~\eqref{eq:pbar} is always satisfied by the following uniform PE condition.

\begin{assum}[Uniform persistent excitation]\label{ass:param}
	There exists $T\in\mathbb{I}_{\geq0}$ such that
	\begin{equation*}
		\Big(\{(x_t,u_t,w_t,p)\}_{t=0}^{K-1},\{(\tilde{x}_t,u_t,\tilde{w}_t,\tilde{p})\}_{t=0}^{K-1}\Big)\in\mathbb{E}_K
	\end{equation*}
	for all $K{\in\,}\mathbb{I}_{\geq T}$, and all trajectories $\{(x_t,u_t,w_t,p)\}_{t=0}^{K-1}\in\mathbb{Z}^K$ and $\{(\tilde{x}_t,u_t,\tilde{w}_t,\tilde{p})\}_{t=0}^{K-1}\in\mathbb{Z}^K$ satisfying~\eqref{eq:sys} for all $t\in\mathbb{I}_{[0,K-1]}$.
\end{assum}
Assumption~\ref{ass:param} essentially imposes that \emph{any} two system trajectories of certain length form a persistently excited trajectory pair.
Robust stability guarantees for joint state and parameter estimation under Assumptions~\ref{ass:IOSS} and~\ref{ass:param} are provided by Corollary~\ref{cor:kappa} (with $\kappa=0$); however, we have the following implications, which are proven below.

\begin{prop}\label{prop:jointIOSS}
	Consider the system~\eqref{eq:sys}. The following statements are equivalent:
		\begin{enumerate}[(a)]
			\item Assumptions~\ref{ass:IOSS} and~\ref{ass:param} hold.
			\item There exists a joint i-IOSS Lyapunov function $G : \mathbb{X}\times\mathbb{X}\times\mathbb{P}\times\mathbb{P}\rightarrow\mathbb{R}_{\geq0}$ such that, for some $\underline{G},\overline{G},Q,R\succ0$ and a constant $\eta\in[0,1)$,
			\begin{align}
				&\left\|\begin{bmatrix} x-\tilde{x}\\ p - \tilde{p}\end{bmatrix}\right\|_{\underline{G}}^2 \leq G(x,\tilde{x},p,\tilde{p}) \leq \left\|\begin{bmatrix} x-\tilde{x}\\ p - \tilde{p}\end{bmatrix}\right\|_{\overline{G}}^2,\label{eq:joint_IOSS_bounds}\\[1ex]
				&G(f(x,u,w,p),f(\tilde{x},u,\tilde{w},\tilde{p}),p,\tilde{p}) \nonumber \\
				&\hspace{3ex} \leq \eta G(x,\tilde{x},p,\tilde{p}) + \|w-\tilde{w}\|_{Q}^2\nonumber  \\
				& \hspace{3ex} \phantom{\leq} \ + \|h(x,u,w,p)-h(\tilde{x},u,\tilde{w},\tilde{p})\|_{R}^2 \label{eq:joint_IOSS_dissip}
			\end{align}
			for all $(x,u,w,p),(\tilde{x},u,\tilde{w},\tilde{p}) \in \mathbb{Z}$.
	\end{enumerate}
\end{prop}

\input{extended_version_prop10.tex}

Proposition~\ref{prop:jointIOSS} essentially implies that state detectability (Assumptions~\ref{ass:IOSS}) and uniform PE of the parameters (Assumption~\ref{ass:param}) is equivalent to uniform detectability (exponential i-IOSS) of the \emph{augmented state} $x_\mathrm{a}^\top=[x^\top,p^\top]$.
Consequently, under these assumptions one could simply consider the augmented state $x_\mathrm{a}$ and apply MHE schemes and theory for state estimation (e.g., \cite{Allan2021a,Knuefer2023,Schiller2023c}).
However, Assumption~\ref{ass:param} is restrictive, usually not satisfied in practice, and its \textit{a priori} verification is generally impossible.
The proposed method from Section~\ref{sec:MHE_nonPE_design}, on the other hand, provides a strict relaxation, since it is applicable in the practically relevant case where the parameters are only rarely (or never) excited (which violates Assumption~\ref{ass:param} and hence implies that the augmented state cannot be uniformly detectable (i.e., exponentially i-IOSS) and no joint i-IOSS Lyapunov function satisfying \eqref{eq:joint_IOSS_bounds} and~\eqref{eq:joint_IOSS_dissip} can exist).

\input{extended_version_verify_PE.tex}

\section{Numerical example}\label{sec:example}

To illustrate our results, we consider the following system
\begin{align*}
	x_1^+ &= x_1 + t_{\Delta}b_1(x_2-a_1x_1-a_2x_1^2-a_3x_1^3) + w_1,\\
	x_2^+ &= x_2 + t_{\Delta}(x_1-x_2+x_3) + w_2,\\
	x_3^+ &= x_3-t_{\Delta}b_2x_2 + w_3,\\
	y &= x_1 + w_4,
\end{align*}
which corresponds to the Euler-discretized modiﬁed Chua's circuit system from~\cite{Yang2015} using the sampling interval $t_{\Delta} = 0.01$ under additional disturbances $w\in\mathbb{R}^4$ and {(noisy)} output measurements.
The parameters are $b_1=12.8$, $b_2 = 19.1$, $a_1=0.6$, $a_2 = -1.1$, $a_3 = 0.45$, which leads to a chaotic behavior of the system.
In the following, we treat $w$ as a uniformly distributed random variable with $|w_i|\leq 10^{-3}, i=1,2,3$ for the process disturbance and $|w_4|\leq 0.1$ for the measurement noise.
We consider the initial condition $x_0=[1,0,-1]^\top$ and assume that $x_t$ evolves in the (known) set $\mathbb{X} = [-1,3]\times[-1,1]\times[-3,3]$.
Furthermore, we consider the case where the exact parameter $a_3=:p$ is unknown but contained in the set $\mathbb{P}=[0.2,0.8]$.
The objective is to compute the state and parameter estimates $\hat{x}_t$ and $\hat{p}_t$ by applying the MHE scheme proposed in Section~\ref{sec:MHE_nonPE} using the initial estimates $\hat{x}_0=[-1,0.1,2]^\top$ and $\hat{p}_0=0.2$.

We construct the i-IOSS Lyapunov function $U$ (Assumption~\ref{ass:IOSS}) and the set $\mathbb{E}_N$ (Definition~\ref{def:obs}) using the methods from \cite[Sec.~IV]{Schiller2023c} and Section~\ref{sec:dIOSS}, respectively.
The computations are carried out in MATLAB using YALMIP~\cite{Loefberg2009} and the semidefinite programming solver MOSEK~\cite{MOSEKApS2019}; details concerning the verification procedure and MHE simulation (including the code) can be found online \cite{Schiller2024d}.
We select constant weighting matrices $W_t = 2P_\mathrm{p}$, $V_t = 100S_\mathrm{p}$, $Q_t = 2(Q_\mathrm{x} + Q_\mathrm{p})$, and $R_t=R_\mathrm{x}+R_\mathrm{p}$ for all $t\in\mathbb{I}_{\geq0}$, the discount factors $\eta_1=0.934$, and $\eta_2=0.9997$, and the horizon length $N=150$. These choices satisfy Assumption~\ref{ass:bounds2} and the contraction conditions~\eqref{eq:contraction_2}--\eqref{eq:contraction_1} (the minimal horizon length for the selected parameters is $N_{\min}=137$). While the relatively high value of the horizon length indicates conservatism in our results, it also seems natural here, as the sampling interval $t_\Delta$ is relatively small compared to the dynamics of the system.
In the following, we consider the MHE scheme presented in Section~\ref{sec:MHE_nonPE} with two different settings: first, without explicit excitation monitoring by naively assuming uniform PE (Assumption~\ref{ass:param}); second, using the proposed excitation-dependent adaptive regularization and selecting the parameter estimate $\hat{p}_t$ in accordance with Remark~\ref{rem:est_p}.
For the latter, we use the method proposed in Section~\ref{sec:dIOSS} and check if $X_t\in\mathbb{E}_{N_t}$ by evaluating $\mathcal{O}_{N_t}(Z_t^*)$ defined in~\eqref{eq:O_T} at the current optimal solution  $Z_t^* = \big(\hat{x}_{t-N_t|t}^*,\hat{p}^*_{|t},\{\hat{w}^*_{j|t}\}_{j=t-N_t}^{t-1}\big)$. If the lower bound $\alpha_t:=\underline{\lambda}(\mathcal{O}_T(Z_t^*))$ satisfies $\alpha_t\geq\alpha$ for the predefined threshold $\alpha=10^{-3}$, we consider that $X_t\in\mathbb{E}_{N_t}$, and $X_t\notin\mathbb{E}_{N_t}$ otherwise.

\begin{figure}
	\flushright
	\includegraphics[width=\columnwidth]{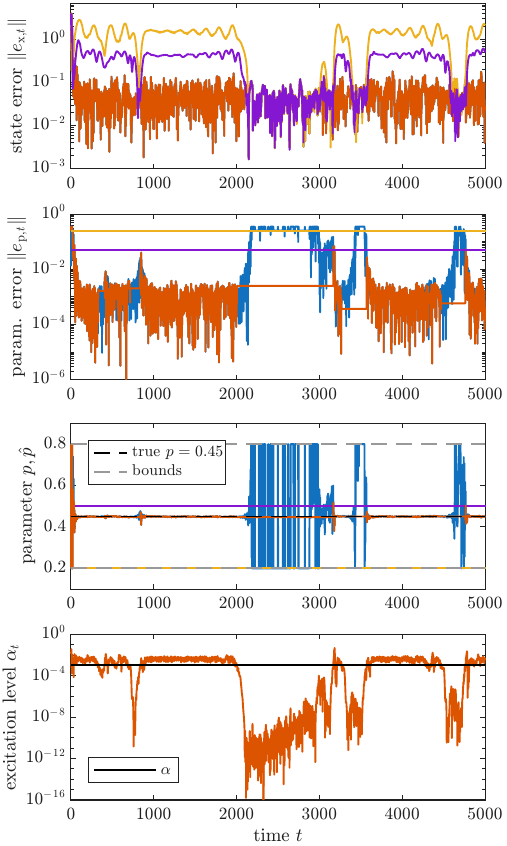}
	\caption{Estimation results of the proposed MHE scheme with adaptive regularization (red) compared to naive MHE without excitation motoring (blue) and two MHE designs for state estimation only, relying on inaccurate (fixed) parameters $\hat{p}=\hat{p}_0=0.2$ (yellow) and $\hat{p}=0.5$ (purple). In the top plot, the red and blue curves coincide.}
	\label{fig}
\end{figure}

The estimation results are shown Figure~\ref{fig}.
Here, we first note that MHE can lead to significant state estimation errors under parametric uncertainties if incorrect (fixed) parameters are used (compare the yellow and purple curves), which highlights the necessity and effectiveness of estimating states and parameters jointly in a combined MHE framework.
In particular, both the naive and adaptive (joint) MHE schemes show rapid convergence of the state and parameter estimation errors, see the blue and red curves. Here, the high accuracy of the parameter estimates at the beginning (see the second and third plot) results from a sufficiently high excitation (compare the bottom plot).
During the simulation, however, there are also phases of weak \mbox{excitation---especially} in the time interval $[2000,3000]$, where the excitation level~$\alpha_t$ is close to zero.
This essentially renders the parameter unobservable, thereby violating Assumption~\ref{ass:param}. As a result, the naive MHE scheme provides poor parameter estimates in this interval, which can be clearly seen from the blue curve in the second and third plots in Figure~\ref{fig}.
In contrast, MHE with the proposed adaptive regularization explicitly takes the excitation level into account and is thus able to efficiently compensate for phases of weak excitation, resulting in significantly better parameter estimation results (see the red curves).

While parameter unobservability must obviously be addressed appropriately to ensure reliable parameter estimates, it has almost no impact on the state estimates (the blue and red curves coincide in the top plot in Figure~\ref{fig}). In fact, the state trajectories produced by the naive and the adaptive MHE schemes are almost identical, which can be attributed to the fact that the parameter has a vanishing influence on the cost function when it is unobservable.
This is also evident in Table~\ref{tab:RMSE}, where we provide the root mean square error (RMSE) for the state and parameter estimates. Specifically, the RMSE of the state estimation errors is equal for both MHE schemes (their difference lies in the order of magnitude of $10^{-15}$), whereas the RMSE of the parameter estimation error for MHE using adaptive regularization is more than six times smaller than the naive MHE scheme without excitation monitoring.

\begin{table}[t]
	\caption{Naive and adaptive MHE for state and parameter estimation.}
	\label{tab:RMSE}
	\begin{threeparttable}
		\setlength\tabcolsep{8pt}
		\begin{tabular*}{\columnwidth}{@{\extracolsep{\fill}}cccc}
			\hline
			\ \ MHE setup & $\mathrm{RMSE}(e_\mathrm{x})$ & $\mathrm{RMSE}(e_\mathrm{p})$ & $\tau_{\mathrm{avrg}}\, [\mathrm{ms}]$ \\
			\hline
			\ \ naive & 0.1707 &  0.1379 & 50.01\\ 
			\ \ adaptive & 0.1707 &  0.0209 & 49.96\\
			\hline
		\end{tabular*}	
		\begin{tablenotes}
			\footnotesize
			\item $\mathrm{RMSE}(e_i){\,:=\,} \sqrt{\frac{1}{N_\mathrm{sim}}\sum_{t=0}^{N_\mathrm{sim}}\|e_{i,t}\|^2}$, $i=\{\mathrm{x},\mathrm{p}\}$, $N_\mathrm{sim}{\,=\,}5000$
		\end{tablenotes}
	\end{threeparttable}
\end{table}

We now compare the average time $\tau_{\mathrm{avrg}}$ required to solve the nonlinear program in~\eqref{eq:MHE} at each time step $t$ for MHE with and without adaptive regularization using a standard laptop.
Interestingly, the last column in Table~\ref{tab:RMSE} shows that the computation times are very similar. This can be attributed to the fact that the additional computation of $\alpha_t$ required for the adaptation mechanism (which mainly consists of matrix operations) saves time in solving the actual optimization problem due to a better conditioning of the cost function compared to naive MHE, especially due to the regularization update~\eqref{eq:pbar}.

\section{Conclusion}

In this paper, we have proposed an MHE scheme to estimate the states and the unknown constant parameters of general uncertain nonlinear discrete-time systems.
The cost function involves an adaptive regularization term that is adjusted according to the current parameter excitation.
We have derived a bound for the state and parameter estimation error that is valid regardless of excitation of the parameter, and in particular also applies if the parameter is never or only rarely excited during operation.
Furthermore, we provided a practical method for online excitation monitoring that is applicable to general nonlinear systems.
The numerical example has shown that the proposed MHE scheme is able to efficiently compensate for phases of weak excitation and ensures reliable estimation results for all times.
An extension of the MHE scheme for time-varying parameters is provided in \cite{Schiller2024a}.
Future work might consider component-wise adaptation mechanisms in MHE depending on component-wise excitation monitoring to further improve the overall estimation performance.

\begin{ack}
	This work was supported by the Deutsche	Forschungsgemeinschaft (DFG, German Research Foundation), Project 426459964.	
\end{ack}

\input{references.bbl}
\appendix
\section{Proofs of Lemmas~\ref{lem:nonPE} and~\ref{lem:PE}}\label{sec:appendix}
\begin{pf*}{Proof of Lemma~\ref{lem:nonPE}.}
	Consider the function~\eqref{eq:candidate}. Since we have that $\hat{x}_t=\hat{x}^*_{t|t}$, boundedness of $V_t$ from \eqref{eq:bound_V2} implies
	\begin{equation}
	\Gamma_1(t,\hat{x}_t,x_t,\hat{p}_t,p)\leq  U(\hat{x}_{t|t}^*,x_{t}) + \overline{\lambda}(\overline{V},\underline{V})\|p-\hat{p}_{t}\|_{\underline{V}}^2. \label{eq:proof_start}
	\end{equation}
	Due to satisfaction of the MHE constraints~\eqref{eq:MHE_1}--\eqref{eq:MHE_con_x}, we can invoke Assumption~\ref{ass:IOSS} (in particular, the dissipation inequality~\eqref{eq:IOSS_dissip}), which lets us conclude that
	\begin{align}
	&U(\hat{x}_{t|t}^*,x_{t})\label{eq:proof_U}\\
	&\leq  \eta_\mathrm{x}^{N_t}U(\hat{x}_{t-N_t|t}^*,x_{t-N_t}) +  \sum_{j={1}}^{N_t}\eta_\mathrm{x}^{j-1}\|\hat{p}^*_{|t}-p\|_{S_\mathrm{x}}^2\nonumber\\
	&\quad + \sum_{j={1}}^{N_t}\eta_\mathrm{x}^{j-1}(\|\hat{w}^*_{t-j|t}-w_{t-j}\|_{Q_\mathrm{x}}^2+\|\hat{y}^*_{t-j|t}-y_{t-j}\|_{R_\mathrm{x}}^2). \nonumber
	\end{align}
	Exploiting that $\|\hat{p}^*_{|t}-p\|_{S_\mathrm{x}}^2\leq \bar{\lambda}(S_\mathrm{x},\underline{V})\|\hat{p}^*_{|t}-p\|^2_{{\underline{V}}}$ and the geometric series, we obtain
	\begin{equation}
		\sum_{j={1}}^{N_t}\eta_\mathrm{x}^{j-1}\|\hat{p}^*_{|t}{-}p\|_{S_\mathrm{x}}^2{\,\leq\;} \bar{\lambda}(S_\mathrm{x},\underline{V})\frac{1-\eta_\mathrm{x}^{N_t}}{1-\eta_\mathrm{x}}\|\hat{p}^*_{|t}{-}p\|^2_{{\underline{V}}}. \label{eq:proof_sum_param}
	\end{equation}
	The upper bound in~\eqref{eq:IOSS_bounds} implies that $U(\hat{x}_{t-N_t|t}^*,x_{t-N_t})\leq \|\hat{x}_{t-N_t|t}^*-{x}_{t-N_t}\|_{\overline{U}}^2$.
	Using Jensen's inequality, we have
	\begin{align}
	&\|\hat{x}^*_{t-N_t|t}-x_{t-N_t}\|^2_{\overline{U}}\leq  2\|\hat{x}^*_{t-N_t|t}-\bar{x}_{t-N_t}\|^2_{\overline{U}}\nonumber \\
	&\hspace{3.5cm} + 2\|\bar{x}_{t-N_t}-x_{t-N_t}\|^2_{\overline{U}}, \phantom{x} \label{eq:proof_triangle_x}\\
	&\|\hat{p}^*_{|t}-p\|^2_{\underline{V}} \leq 2\|\hat{p}^*_{|t}-\bar{p}_{t-N_t}\|^2_{\underline{V}} + 2\|\bar{p}_{t-N_t}-p\|^2_{\underline{V}}, 
	\label{eq:proof_triangle_p}\\
	&\|\hat{w}^*_{t-j|t}-w_{t-j}\|^2_{Q_\mathrm{x}}\leq 2\|\hat{w}^*_{t-j|t}\|^2_{Q_\mathrm{x}} + 2\|w_{t-j}\|^2_{Q_\mathrm{x}} \label{eq:proof_triangle_w}
	\end{align}
	for all $j\in\mathbb{I}_{[1,N_t]}$. Applying~\eqref{eq:proof_U}--\eqref{eq:proof_triangle_p} to \eqref{eq:proof_start} leads to
	\begin{align*}
		&\Gamma_1(t,\hat{x}_t,x_t,\hat{p}_t,p)\nonumber\\
		&\leq \eta_\mathrm{x}^{N_t}(2\|\hat{x}_{t-N_t|t}^*-\bar{x}_{t-N_t}\|_{\overline{U}}^2+2\|\bar{x}_{t-N_t}-x_{t-N_t}\|_{\overline{U}}^2)\nonumber\\
		&\quad + c_1(N_t)(2\|\hat{p}^*_{|t}-\bar{p}_{t-N_t}\|^2_{\underline{V}} + 2\|\bar{p}_{t-N_t}-p\|^2_{\underline{V}})\nonumber\\
		&\quad  + \sum_{j={1}}^{N_t}\eta_\mathrm{x}^{j-1}(2\|\hat{w}^*_{t-j|t}\|_{Q_\mathrm{x}}^2+2\|w_{t-j}\|_{Q_\mathrm{x}}^2\nonumber\\
		&\hspace{2.1cm} + \|\hat{y}^*_{t-j|t}-y_{t-j}\|_{R_\mathrm{x}}^2),
	\end{align*}
	where we have used the definition of $c_1(s)$ from~\eqref{eq:c1_s}.
	Using that $\eta_1^{N_t-N} \geq 1$ and $c_1(N_t)> 1$, we can invoke the cost function~\eqref{eq:MHE_objective} due to the facts that $\eta_\mathrm{x}^s \leq \gamma(s)$ for all $s\geq 0$, $\eta_\mathrm{x}\leq\eta_2$, and $2\overline{U}\preceq W_t \preceq \overline{W}$, $2\underline{V}\preceq V_t \preceq \overline{V}$, $2Q_\mathrm{x} \preceq Q_t \preceq \overline{Q}$, $R_\mathrm{x} \preceq R_t$ for all $t\in\mathbb{I}_{\geq0}$ by Assumption~\ref{ass:bounds2}, which yields
	\begin{align}
		&\Gamma_1(t,\hat{x}_t,x_t,\hat{p}_t,p)\nonumber\\
		&\leq  \eta_{\mathrm{1}}^{-N}c_{1}(N_t)\label{eq:proof_cost}\\
		&\quad 	\cdot\Big(\eta_\mathrm{x}^{N_t}(\|\bar{x}_{t-N_t}-x_{t-N_t}\|_{\overline{W}}^2)\nonumber + \eta_1^{N_t}\|\bar{p}_{t-N_t}-p\|^2_{\overline{V}}\nonumber\\
		&\qquad + \sum_{j={1}}^{N_t}\eta_2^{j-1}\|w_{t-j}\|_{\overline{Q}}^2 + 	{J_t(\hat{x}_{t-N_t|t}^*,\hat{p}_{|t}^*,\hat{w}_{\cdot|t}^*,\hat{y}_{\cdot|t}^*)}\Big).\nonumber
	\end{align}
	Using optimality and boundedness of $W_t,V_t,Q_t$ yields
	\begin{align}
	&\ J_t(\hat{x}_{t-N_t|t}^*,\hat{p}_{|t}^*,\hat{w}_{\cdot|t}^*,\hat{y}_{\cdot|t}^*) \leq J_t(x_{t-N_t},p,w_{\cdot|t},y_{\cdot|t})\nonumber\\
	&\leq \gamma(N_t)\|x_{t-N_t}-\bar{x}_{t-N_t}\|_{\overline{W}}^2 +  {\eta}_1^{N_t}\|p-\bar{p}_{t-N_t}\|_{\overline{V}}^2\nonumber\\
	&\quad + \sum_{j=1}^{N_t}{\eta}_2^{j-1}\|{w}_{t-j}\|_{\overline{Q}}^2. \label{eq:proof_optimality}
	\end{align}
	Combining \eqref{eq:proof_cost} and \eqref{eq:proof_optimality} yields \eqref{eq:proof_case_1}, which hence concludes this proof. \qed
\end{pf*}

\begin{pf*}{Proof of Lemma~\ref{lem:PE}.}
	We start by following the same arguments as in the beginning of the proof of Lemma~\ref{lem:nonPE} (based on the fact that the optimal estimated trajectory is a trajectory of the \mbox{i-IOSS} system~\eqref{eq:sys} by invoking the MHE constraints~\eqref{eq:MHE_1}--\eqref{eq:MHE_con_x}). This allows us to exploit \eqref{eq:proof_U} and \eqref{eq:proof_sum_param} with $\underline{V}$ replaced by $S_\mathrm{p}$, leading to
	\begin{align*}
		&\Gamma_2(\hat{x}_{t},x_t,\hat{p}_{t},p) =  U(\hat{x}_{t},x_t) + 	c\|\hat{p}_{t}-p\|^2_{\overline{V}}\\
		&\leq \eta_{\mathrm{x}}^N U(\hat{x}_{t-N|t}^*,{x}_{t-N}) + 	c_2(c,N)\|\hat{p}_{|t}^*-p\|^2_{S_\mathrm{p}}\\
		&\quad + 	\sum_{j={1}}^{N}\eta_\mathrm{x}^{j-1}(\|\hat{w}_{t-j|t}^*-w_{t-j}\|_{Q_{\mathrm{x}}}^2+\|\hat{y}_{t-j|t}^*-y_{t-j}\|_{R_\mathrm{x}}^2)
	\end{align*}	
	where we have used $c_2(c,N)$ from~\eqref{eq:c}. In the following, we drop the arguments of $c_2$ for the sake of brevity.
	Since $X_t\in\mathbb{E}_N$, it follows that
	\begin{align*}
		& \|\hat{p}_{|t}^*-p\|^2_{S_{\mathrm{p}}} \nonumber\\	&\leq\eta_\mathrm{p}^N\|\hat{x}_{t-N|t}^*-{x}_{t-N}\|^2_{P_\mathrm{p}} \\
		&\quad + \sum_{j={1}}^N\eta_\mathrm{p}^{j-1}(\|\hat{w}_{t-j|t}^*-{w}_{t-j}\|^2_{Q_{\mathrm{p}}} + 	\|\hat{y}_{t-j|t}^*-{y}_{t-j}\|^2_{R_\mathrm{p}}).
	\end{align*}
	Using the bound on $U$ together with the definition of $\gamma(s)$ from \eqref{eq:gamma} and the fact that $c_2(c,s)\geq1$ for all $s>0$ due to $c\geq1$ and \eqref{eq:bound_V2}, we obtain
	\begin{align*}
		&\Gamma_2(\hat{x}_{t},x_t,\hat{p}_{t},p)\\
		&\leq c_2\Big(\gamma(N)\|\hat{x}_{t-N|t}^*-{x}_{t-N}\|^2_{\overline{U}}\\
		&\quad + 	\sum_{j={1}}^{N}\tilde{\eta}^{j-1}(\|\hat{w}_{t-j|t}^*-w_{t-j}\|_{\tilde{Q}}^2+\|\hat{y}_{t-j|t}^*-y_{t-j}\|_{\tilde{R}}^2)\Big),
	\end{align*}
	where $\tilde{\eta}:=\max\{\eta_{\mathrm{x}},\eta_{\mathrm{p}}\}$, $\tilde{Q} = Q_{\mathrm{x}}+Q_\mathrm{p}$, and $\tilde{R} := R_{\mathrm{x}}+R_\mathrm{p}$.
	Application of~\eqref{eq:proof_triangle_x} and \eqref{eq:proof_triangle_w} with $Q_\mathrm{x}$ replaced by $\tilde{Q}$ together with the definition of $J$ from~\eqref{eq:MHE_objective} and Assumption~\ref{ass:bounds2} leads to
	\begin{align*}
		&\Gamma_2(\hat{x}_{t},x_t,\hat{p}_{t},p)\\
		&\leq c_2\Big(\gamma(N) \|\bar{x}_{t-N}-{x}_{t-N}\|^2_{\overline{W}}\\
		&\qquad + \sum_{j={1}}^{N}{\eta}_2^{j-1}\|w_{t-j}\|_{\overline{Q}}^2 + {J_t(\hat{x}_{t-N_t|t}^*,\hat{p}_{|t}^*,\hat{w}_{\cdot|t}^*,\hat{y}_{\cdot|t}^*)}\Big).
	\end{align*}
	By optimality, the first inequality in \eqref{eq:proof_optimality} holds, which leads to
	\begin{align*}
		&\Gamma_2(\hat{x}_{t},x_t,\hat{p}_{t},p)\\
		&\leq 2c_2\gamma(N) \|\bar{x}_{t-N}-{x}_{t-N}\|^2_{\overline{W}} \\
		&\quad + {c_2}\eta_1^{N} \|\bar{p}_{t-N}-p\|_{V_{t-N}}^2 + 2{c_{2}}\sum_{j={1}}^{N}{\eta}_2^{j-1}\|w_{t-j}\|_{\overline{Q}}^2.\\[-3.312ex]
	\end{align*}
	Application of $\|\bar{x}_{t-N}-{x}_{t-N}\|^2_{\overline{W}} \leq \overline{\lambda}(\overline{W},\underline{U})\|\bar{x}_{t-N}-{x}_{t-N}\|^2_{\underline{U}}$ together with the definitions of $\Gamma_1$ from~\eqref{eq:candidate} and $\mu$ from~\eqref{eq:mu} yields~\eqref{eq:lem_PE2}, which finishes this proof. \qed
\end{pf*}

\end{document}

%% file: extended_version_prop10.tex
\vspace{-0.1ex}
\begin{pf*}{Proof.}
	Consider some $K\in\mathbb{I}_{\geq1}$. Let the two sequences $\{(x_t,u_t,w_t,p)\}_{t=0}^{K-1}\in\mathbb{Z}^K$ and $\{(\tilde{x}_t,u_t,\tilde{w}_t,\tilde{p})\}_{t=0}^{K-1}\in\mathbb{Z}^K$ satisfy \eqref{eq:sys} for all $t\in\mathbb{I}_{[0,K-1]}$. Define the corresponding outputs $y_t=h(x_t,u_t,w_t,p)$ and $\tilde{y}_t=h(\tilde{x}_t,u_t,\tilde{w}_t,\tilde{p})$, $t\in\mathbb{I}_{[0,K-1]}$.
	For the sake of conciseness, define $\Delta x_t := x_t-\tilde{x}_t$ for $t\in\mathbb{I}_{[0,K]}$, $\Delta w_t := w_t-\tilde{w}_t$ and $\Delta y_t := y_t-\tilde{y}_t$ for $t\in\mathbb{I}_{[0,K-1]}$, and $\Delta p := p-\tilde{p}$.
	
	We start with $(a)\Rightarrow(b)$.
	Assumption~\ref{ass:IOSS} implies (by application of~\eqref{eq:IOSS_dissip}, \eqref{eq:IOSS_bounds}, and the geometric series) the following bound:
	\begin{align}
		&\|\Delta x_t\|_{\underline{U}}^2 + \|\Delta p\|_{S_\mathrm{x}}^2 \nonumber\\
		&\leq \eta_\mathrm{x}^t \|\Delta x_0\|_{\overline{U}}^2 + \left(\frac{1}{1-\eta_{\mathrm{x}}}+1\right)\|\Delta p\|_{S_\mathrm{x}}^2\nonumber\\
		&\quad + \sum_{j=1}^t  \eta_\mathrm{x}^{j-1}(\|\Delta w_{t-j}\|_{Q_\mathrm{x}}^2 + \|\Delta y_{t-j}\|_{R_\mathrm{x}}^2)\label{eq:proof_IOSS}
	\end{align}
	for all $t\in\mathbb{I}_{[0,K]}$, where we have added $\|\Delta p\|_{S_\mathrm{x}}^2$ to both sides.
	We make a case distinction and first consider $t\in\mathbb{I}_{[T,K]}$.
	Application of $\|\Delta p\|_{S_\mathrm{x}}^2 \leq \overline{\lambda}(S_\mathrm{x},S_\mathrm{p})\|\Delta p\|_{S_\mathrm{p}}^2$ and Assumption~\ref{ass:param} leads to
	\begin{align}
		&\|\Delta x_t\|_{\underline{U}}^2  + \|\Delta p\|_{S_\mathrm{x}}^2 \leq \tilde{\eta}^t\|\Delta x_0\|^2_{\tilde{P}_1} \notag\\
		& + \sum_{j=1}^t  \tilde{\eta}^{j-1}(\|\Delta w_{t-j}\|_{\tilde{Q}}^2 + \|\Delta y_{t-j}\|_{\tilde{R}}^2), \label{eq:proof_W_1}
	\end{align}
	where we have used the definitions $\tilde{\eta} := \max\{\eta_{\mathrm{x}},\eta_\mathrm{p}\}$, $\tilde{P}_1:=\overline{U} + c_1P_\mathrm{p}$, $\tilde{Q}:= Q_\mathrm{x} + c_1Q_\mathrm{p}$, and $\tilde{R}:= R_\mathrm{x} + c_1R_\mathrm{p}$ with $c_1:=\left(\frac{1}{1-\eta_{\mathrm{x}}}+1\right) \overline{\lambda}(S_\mathrm{x},S_\mathrm{p})$.
	Now, recall that~\eqref{eq:proof_IOSS} also applies for $t\in\mathbb{I}_{[0,T-1]}$. Using the fact that $1\leq\tilde{\eta}^{1-T}\tilde{\eta}^t$ for all $t\in\mathbb{I}_{[0,T-1]}$, one can verify that
	\begin{align}
		&c_2(\|\Delta x_t\|^2 + \|\Delta p\|^2) \nonumber\\
		&\leq c_3\tilde{\eta}^t\left( \|\Delta x_0\|^2 + \|\Delta p\|^2\right)\nonumber\\
		&\quad + \sum_{j=1}^t  \tilde{\eta}^{j-1}(\|\Delta w_{t-j}\|_{\tilde{Q}}^2 + \|\Delta y_{t-j}\|_{\tilde{R}}^2)\label{eq:proof_W_IOSS}
	\end{align}
	for all $t\in\mathbb{I}_{[0,K]}$, where $c_2 := \min\{\underline{\lambda}(\underline{U}),\underline{\lambda}(S_\mathrm{x})\}$, $c_3:=\max\left\{\overline{\lambda}(\tilde{P}_1),\overline{\lambda}(S_\mathrm{x})\tilde{\eta}^{1-T}\left(\frac{1}{1-\eta_{\mathrm{x}}}+1\right)\right\}$.
	Consider the augmented states $x_{\mathrm{a},t}^\top = \left[ x_t^\top, p^\top \right]$ and $\tilde{x}_{\mathrm{a},t}^\top = \left[ \tilde{x}_{t}^\top, \tilde{p}^\top \right]$, which evolve according to the augmented system dynamics
	\begin{equation}
		x_\mathrm{a}^+ = f_\mathrm{a}(x_\mathrm{a},u,w) =
		\begin{bmatrix} f(x,u,w,p)\\ p \end{bmatrix}.\label{eq:proof_W_sys}
	\end{equation}
	By satisfaction of~\eqref{eq:proof_W_IOSS} and the fact that $\|\Delta x_t\|^2 + \|\Delta p\|^2 =  \|x_{\mathrm{a},t} - \tilde{x}_{\mathrm{a},t}\|^2$, we observe that the system~\eqref{eq:proof_W_sys} is exponentially \mbox{i-IOSS} \cite[Def.~4.5]{Rawlings2017} with respect to the outputs $y_\mathrm{a} = h_\mathrm{a}(x_\mathrm{a},u,w) := h(x,u,w,p)$.
	Existence of an \mbox{i-IOSS} Lyapunov function $G(\cdot)$ and suitable matrices $\underline{G},\overline{G},Q,R\succ0$ satisfying~\eqref{eq:joint_IOSS_bounds} and \eqref{eq:joint_IOSS_dissip} follows by a straightforward extension of the converse Lyapunov theorem from \cite{Allan2021}.
	
	It remains to show $(b)\Rightarrow(a)$.  Application of~\eqref{eq:joint_IOSS_dissip} and \eqref{eq:joint_IOSS_bounds} yields
	\begin{align}
		&\underline{\lambda}(\underline{G})(\|\Delta x_t\|^2 + \|\Delta p\|^2) \leq \overline{\lambda}(\overline{G})\eta^t(\|\Delta x_0\|^2 + \|\Delta p\|^2)\nonumber\\
		& \hspace{10ex} + \sum_{j=1}^{t}\eta^{j-1}(\|\Delta w_{t-j}\|^2_Q + \|\Delta y_{t-j}\|^2_R)\label{eq:proof_W_2}
	\end{align}
	for all $t\in\mathbb{I}_{[0,K]}$.
	Using that $\|\Delta p\|\geq0$, we obtain
	\begin{align*}
		&\underline{\lambda}(\underline{G})\|\Delta x_t\|^2 \leq \overline{\lambda}(\overline{G})\eta^t\|\Delta x_0\|^2\\
		&\quad + \sum_{j=1}^{t}\eta^{j-1}(\overline{\lambda}(\overline{G})\|\Delta p\|^2 + \|\Delta w_{t-j}\|^2_Q + \|\Delta y_{t-j}\|^2_R),
	\end{align*}
	which is an (exponential) i-IOSS bound for the system~\eqref{eq:sys} considering $p$ as an additional constant input. Existence of an i-IOSS Lyapunov function $U(x,\tilde{x})$ and matrices $\underline{U},\overline{U},Q_\mathrm{x},R_\mathrm{x}\succ0$ and $\eta_\mathrm{x}\in[0,1)$ satisfying Assumption~\ref{ass:IOSS} follows by a straightforward extension of the converse Lyapunov theorem from~\cite{Allan2021}.
	Now fix some $T\in\mathbb{I}_{\geq1}$ and consider some $K\in\mathbb{I}_{\geq T}$.
	From~\eqref{eq:proof_W_2} with $t=K$ and the facts that $\eta^K\leq \eta^{T}$ and $\|\Delta x_K\|^2\geq 0$, we obtain
	\begin{align*}
		&(\underline{\lambda}(\underline{G})- \overline{\lambda}(\overline{G})\eta^{T})\|\Delta p\|^2 \leq \overline{\lambda}(\overline{G})\eta^K\|\Delta x_0\|^2\\
		&\quad + \sum_{j=1}^{K}\eta^{j-1}(\|\Delta w_{K-j}\|^2_Q + \|\Delta y_{K-j}\|^2_R)
	\end{align*}
	for all $K\in\mathbb{I}_{\geq T}$. Since there always exists $T\in\mathbb{I}_{\geq 1}$ such that $(\underline{\lambda}(\underline{G})-\overline{\lambda}(\overline{G})\eta^{T})>0$, Assumption~\ref{ass:param} is satisfied, which finishes this proof. \qed
\end{pf*}

%% file: extended_version_verify_PE.tex
\section{Verifying persistent excitation}
\label{sec:dIOSS}

The MHE scheme from Section~\ref{sec:MHE_nonPE} requires detecting whether the PE condition in the update law~\eqref{eq:pbar} at a given time $t\in\mathbb{I}_{\geq N}$ is satisfied or not.
To this end, we propose a sufficient condition that can be verified online.

In the following, we use the definitions $z := (x,u,w,p)$ and $\tilde{z} := (\tilde{x},u,\tilde{w},\tilde{p})$ for any $(x,u,w,p),(\tilde{x},u,\tilde{w},\tilde{p})\in\mathbb{Z}$.
In the remainder of this section, we assume that $f$ and $h$ are at least twice continuously differentiable in all of its arguments.
Let $z_s(s) := z + s(\tilde{z}-z)$ for $s\in[0,1]$ and
\begin{align}
	&\textstyle A(z,\tilde{z}) := \int_0^1\frac{\partial f}{\partial x}(z_s(s))ds,
	\hspace{2.6ex} \textstyle C(z,\tilde{z}) := \int_0^1\frac{\partial h}{\partial x}(z_s(s))ds,\nonumber \\
	&\textstyle B(z,\tilde{z}) := \int_0^1\frac{\partial f}{\partial w}(z_s(s))ds,
	\hspace{2ex} \textstyle D(z,\tilde{z}) := \int_0^1\frac{\partial h}{\partial w}(z_s(s))ds,\nonumber \\
	&\textstyle E(z,\tilde{z}) := \int_0^1\frac{\partial f}{\partial p}(z_s(s))ds,
	\hspace{2.4ex} \textstyle F(z,\tilde{z}) := \int_0^1\frac{\partial h}{\partial p}(z_s(s))ds\label{eq:lin}
\end{align}
for all $z,\tilde{z}\in\mathbb{Z}$. In the following, we require some boundedness properties of the terms in~\eqref{eq:lin}.

\begin{assum}[Bounded linearizations]\label{ass:bounds_MVT}
	There exist constants $\bar{B},\bar{C},\bar{D}\geq0$ such that $\|B(z,\tilde{z})\|\leq \bar{B}$, $\|C(z,\tilde{z})\|\leq \bar{C}$, $\|D(z,\tilde{z})\|\leq \bar{D}$ for all $z,\tilde{z}\in\mathbb{Z}$.
\end{assum}

Assumption \ref{ass:bounds_MVT} is naturally satisfied for special classes of systems (e.g., with additive $w$ and $h$ linear in $x$, which renders $B,C,D$ constant) or generally if $\mathbb{Z}$ is compact.

\begin{assum}[State detectability]\label{ass:obs}
	There exists a mapping $L:\mathbb{Z}\times\mathbb{Z}\rightarrow\mathbb{R}^{n\times p}$, a symmetric matrix $P\succ0$, and constant $\eta\in(0,1)$ such that
	\begin{equation}\label{eq:PHI}
		\Phi(z,\tilde{z}) = A(z,\tilde{z}) + L(z,\tilde{z})C(z,\tilde{z})
	\end{equation}
	satisfies
	\begin{equation}\label{eq:PHI_eta}
		\Phi(z,\tilde{z})^\top P\Phi(z,\tilde{z}) \preceq \eta P
	\end{equation}
	for all $z,\tilde{z}\in\mathbb{Z}$.
	Furthermore, there exists $\bar{L}>0$ such that $\|L(z,\tilde{z})\|\leq \bar{L}$ for all $z,\tilde{z}\in\mathbb{Z}$.
\end{assum}

\begin{rem}
	Assumption~\ref{ass:obs} is motivated by linear systems theory, where detectability is equivalent to the existence of an output injection term which renders the error system asymptotically stable.
	For any fixed $\eta\in[0,1)$, by using the Schur complement and the definition $\mathcal{Y}(z,\tilde{z}) := P L(z,\tilde{z})$, condition~\eqref{eq:PHI_eta} can be transformed into an infinite set of LMIs (linear in the decision variables $P$ and $\mathcal{Y}$).
	Then, these may be solved under a suitable parameterization of $\mathcal{Y}$ (e.g., polynomial in $z,\tilde{z}$) using a finite set of LMIs and standard convex analysis tools based on semidefinite programming (SDP), e.g., by applying sum-of-squares relaxations \cite{Parrilo2003}, by embedding the nonlinear behavior in an LPV model \cite{Sadeghzadeh2023}, or by suitably gridding the state space and verifying~\eqref{eq:PHI_eta} on the grid points (assuming compactness of $\mathbb{Z}$).
	We also want to emphasize that $L$ does not need to be constant as it is usually required in the context of (adaptive) observer design in order to be able to perform the observer update recursions, cf., e.g.,~\cite{Ibrir2018}.
	Instead, the additional degree of freedom resulting from the fact that $L$ may depend on both $z$ and $\tilde{z}$ can be used, e.g., to compensate for nonlinear terms in~$A(z,\tilde{z})$ and/or $C(z,\tilde{z})$ from \eqref{eq:lin}.
	Furthermore, if $L(z,\tilde{z})$ can be chosen such that $\Phi(z,\tilde{z})$ in~\eqref{eq:PHI} becomes constant, the condition~\eqref{eq:PHI_eta} can be drastically simplified (to one single LMI).
\end{rem}

Consider the trajectory pair
\begin{equation}
	\left(\{(x_t,u_t,w_t,p)\}_{t=0}^{T-1},\{(\tilde{x}_t,u_t,\tilde{w}_t,\tilde{p})\}_{t=0}^{T-1}\right) \in \mathbb{Z}^T\times\mathbb{Z}^T \label{eq:traj}
\end{equation}
for some $T\in\mathbb{I}_{\geq0}$, where $x_{t+1}=f(x_t,u_t,w_t,p)$ and $\tilde{x}_{t+1}=f(\tilde{x}_t,u_t,\tilde{w}_t,\tilde{p})$ for all $t\in\mathbb{I}_{[0,T-1]}$.
For the sake of brevity, define
\begin{equation*}
	z_t := (x_t,u_t,w_t,p), \ \tilde{z}_t := (\tilde{x}_t,u_t,\tilde{w}_t,\tilde{p})
\end{equation*}
for all $t\in\mathbb{I}_{[0,T-1]}$, and
\begin{align}
	Z &:= \left(x_0,p,\{{u}_t\}_{t=0}^{T-1},\{w_t\}_{t=0}^{T-1}\right), \label{eq:Z}\\
	\tilde{Z} &:= \left(\tilde{x}_0,\tilde{p},\{{u}_t\}_{t=0}^{T-1},\{\tilde{w}_t\}_{t=0}^{T-1}\right). \label{eq:Z_tilde}
\end{align}

The following result provides a sufficient condition for the trajectory pair~\eqref{eq:traj} to be an element of the set $\mathbb{E}_T$ satisfying Definition~\ref{def:obs} based on matrix recursions.

\begin{prop}\label{prop:C_PE}
	Let Assumptions \ref{ass:bounds_MVT} and \ref{ass:obs} hold.
	Suppose that for some fixed $T\in\mathbb{I}_{\geq1}$ and $\alpha>0$, the trajectories~\eqref{eq:traj} satisfy
	\begin{align}
		\mathcal{C}_T(Z,\tilde{Z}) {\,:=\hspace{-0.3ex}} \sum_{t=0}^{T-1}\mu^{T-1-t}\overline{Y}_t(z_t,\tilde{z}_t)^\top \overline{Y}_t(z_t,\tilde{z}_t){\succ\,} \alpha I_o \label{eq:cond_C}
	\end{align}
	with $\overline{Y}_t(z_t,\tilde{z}_t) {:=\,} C(z_t,\tilde{z}_t)Y_t{+\,}F(z_t,\tilde{z}_t)$, where $Y_t$ {satisfies}
	\begin{align}
		&Y_{t+1} = \Phi(z_t,\tilde{z}_t)Y_t + E(z_t,\tilde{z}_t) + L(z_t,\tilde{z}_t)F(z_t,\tilde{z}_t) \label{eq:Y_def}
	\end{align}
	for all $t\in\mathbb{I}_{[0,T-1]}$ and $Y_0  = 0_{n\times o}$.
	Then, there exist $Q_\mathrm{p},S_\mathrm{p},R_\mathrm{p}\succ0$, and $\eta_\mathrm{p}\in[0,1)$ such that the trajectory pair~\eqref{eq:traj} is an element of the set $\mathbb{E}_T$ (Definition~\ref{def:obs}) with $P_\mathrm{p}=P$ from Assumption~\ref{ass:obs}.
\end{prop}

\begin{pf*}{Proof.}
	Consider the trajectory pair~\eqref{eq:traj} and the outputs $y_t=h(x_t,u_t,w_t,p)$ and $\tilde{y}_t=h(\tilde{x}_t,u_t,\tilde{w}_t,\tilde{p})$, $t\in\mathbb{I}_{[0,T-1]}$.
	Using the mean-value theorem and the definitions from~\eqref{eq:lin}, we have that
	\begin{align*}
		x_{t+1}-\tilde{x}_{t+1} &= A(z_t,\tilde{z}_t)(x_t-\tilde{x}_t) + B(z_t,\tilde{z}_t)(w_t-\tilde{w}_t)\\
		&\quad + E(z_t,\tilde{z}_t)(p-\tilde{p})
	\end{align*}
	and
	\begin{align}
		y_t-\tilde{y}_t &= C(z_t,\tilde{z}_t)(x_t-\tilde{x}_t) + D(z_t,\tilde{z}_t)(w_t-\tilde{w}_t)\nonumber\\
		&\quad + F(z_t,\tilde{z}_t)(p-\tilde{p}) \label{eq:proof_y_div}
	\end{align}
	for all $t\in\mathbb{I}_{[0,T-1]}$.
	Now consider the transformed coordinates $\zeta_t := x_t - Y_tp$ and $\tilde{\zeta}_t := \tilde{x}_t - Y_t\tilde{p}$, $t\in\mathbb{I}_{[0,T-1]}$, where $Y_t$ is from~\eqref{eq:Y_def}. Since
	\begin{equation}\label{eq:zeta}
		\zeta_t - \tilde{\zeta}_t = x_t - \tilde{x}_t -  Y_t(p-\tilde{p}),
	\end{equation}
	we obtain
	\begin{align*}
		&\zeta_{t+1}-\tilde{\zeta}_{t+1} = x_{t+1} - \tilde{x}_{t+1} - Y_{t+1}(p - \tilde{p})\\
		&= A(z_t,\tilde{z}_t)(x_t-\tilde{x}_t) + B(z_t,\tilde{z}_t)(w_t-\tilde{w}_t) \\
		&\ \ + E(z_t,\tilde{z}_t)(p-\tilde{p})\\
		& \ \ - (\Phi(z_t,\tilde{z}_t)Y_t + E(z_t,\tilde{z}_t) + L(z_t,\tilde{z}_t)F(z_t,\tilde{z}_t))(p-\tilde{p}).
	\end{align*}
	To the right-hand side of the previous equation, we add
	\begin{align}
		0&=L(z_t,\tilde{z}_t)(y_t-\tilde{y}_t-(y_t-\tilde{y}_t))\nonumber\\
		&= L(z_t,\tilde{z}_t)(C(z_t,\tilde{z}_t)(x_t-\tilde{x}_t) + D(z_t,\tilde{z}_t)(w_t-\tilde{w}_t)\nonumber\\
		&\quad +F(z_t,\tilde{z}_t)(p-\tilde{p})) - L(z_t,\tilde{z}_t)(y_t-\tilde{y}_t)\label{eq:proof_L_add}
	\end{align}
	with $L$ from Assumption~\ref{ass:obs}.
	Using the definitions of $\zeta_t,\tilde{\zeta}_t$, and $\Phi$ from~\eqref{eq:PHI}, we obtain
	\begin{align}
		\zeta_{t+1}-\tilde{\zeta}_{t+1} =&\ \Phi(z_t,\tilde{z}_t)(\zeta_t-\tilde{\zeta}_t)\nonumber \\
		& + (B(z_t,\tilde{z}_t){\,+\,}L(z_t,\tilde{z}_t)D(z_t,\tilde{z}_t))(w_t{\,-\,}\tilde{w}_t) \nonumber\\
		& - L(z_t,\tilde{z}_t)(y_t-\tilde{y}_t).\label{eq:proof_start_iIOSS}
	\end{align}
	Applying the norm $\|\cdot\|_P=\sqrt{\|\cdot\|_P^2}$ to both sides and using the triangle inequality leads to
	\begin{align*}
		&\|\zeta_{t+1}-\tilde{\zeta}_{t+1}\|_P \\
		&\leq \|\Phi(z_t,\tilde{z}_t)(\zeta_t-\tilde{\zeta}_t)\|_P \\
		&\quad + \|(B(z_t,\tilde{z}_t)+L(z_t,\tilde{z}_t)D(z_t,\tilde{z}_t))(w_t-\tilde{w}_t)\|_P\\
		&\quad + \|L(z_t,\tilde{z}_t)(y_t-\tilde{y}_t)\|_P.
	\end{align*}
	Now, we square both sides, use the fact that for any $\epsilon>0$, $(a+b)^2\leq (1+\epsilon)a^2 + \frac{1+\epsilon}{\epsilon} b^2$ for all $a,b\geq0$ by Young's inequality, apply Assumption~\ref{ass:obs}, and exploit that $(\sum_{i=1}^n a_i)^2 \leq n \sum_{i=1}^n a_i^2$ for any $n\in\mathbb{I}_{\geq0}$ and $a_i\geq0$, $i\in\mathbb{I}_{[1,n]}$, by Jensen's inequality. This results in
	\begin{align}
		&\|\zeta_{t+1}-\tilde{\zeta}_{t+1}\|_P^2 \nonumber\\
		&\leq (1+\epsilon)\eta\|\zeta_t-\tilde{\zeta}_t\|_P^2 \nonumber\\
		& \ + \frac{2(1{\,+\,}\epsilon)}{\epsilon}(\|(B(z_t,\tilde{z}_t){\,+\,}L(z_t,\tilde{z}_t)D(z_t,\tilde{z}_t))(w_t{\,-\,}\tilde{w}_t)\|_P^2\nonumber\\
		& \qquad + \|L(z_t,\tilde{z}_t)(y_t-\tilde{y}_t)\|_P^2).\label{eq:proof_P}
	\end{align}
	Now consider the recursion
	\begin{equation}
		S_{t+1} = \mu S_{t} + \overline{Y}_t(z_t,\tilde{z}_t)^\top\overline{Y}_t(z_t,\tilde{z}_t), \ t\in\mathbb{I}_{0,T-1},\label{eq:S_def}
	\end{equation}
	with $S_0 = 0_{o\times o}$, $\overline{Y}_t$ from~\eqref{eq:cond_C}, and some $\mu\in[0,1)$ that will be specified below.
	Using~\eqref{eq:S_def}, we can write that
	\begin{align}
		\|p-\tilde{p}\|^2_{S_{t+1}} =&\ \mu\|p-\tilde{p}\|^2_{S_t} + \|\overline{Y}_t(z_t,\tilde{z}_t)(p-\tilde{p})\|^2. \label{eq:proof_St}
	\end{align}
	Furthermore, by the definition of $\overline{Y}_t$, the transformation~\eqref{eq:zeta}, and \eqref{eq:proof_y_div}, it follows that
	\begin{align}
		&\|\overline{Y}_t(z_t,\tilde{z}_t)(p-\tilde{p})\|^2\nonumber\\
		&= \|y_t-\tilde{y}_t - D(z_t,\tilde{z}_t)(w_t-\tilde{w}_t) - C(z_t,\tilde{z}_t)(\zeta_t-\tilde{\zeta}_t)\|^2\nonumber\\
		&\leq 3\|y_t-\tilde{y}_t\|^2 + 3\|D(z_t,\tilde{z}_t)(w_t-\tilde{w}_t)\|^2\nonumber\\
		&\quad + 3\frac{\bar{C}^2}{\underline{\lambda}(P)}\|\zeta_t-\tilde{\zeta}_t\|^2_P,
		\label{eq:proof_CY_bound}
	\end{align}
	where the last step followed by applying Jensen's inequality, Assumption~\ref{ass:bounds_MVT}, and $P$ from Assumption~\ref{ass:obs}.
	Now consider the function $W(t,{\zeta},\tilde{\zeta},p,\tilde{p}):= \|\zeta-\tilde{\zeta}\|_P^2 + \gamma\|p-\tilde{p}\|_{S_t}^2$ $t\in\mathbb{I}_{[0,T]}$ for some $\gamma>0$.
	We choose the constants $\mu,\epsilon,\gamma$ introduced above such that
	\begin{equation}
		\mu = (1+\epsilon)\eta + \gamma {3\bar{C}^2}/{\underline{\lambda}(P)} < 1. \label{eq:proof_mu}
	\end{equation}
	Using \eqref{eq:proof_P}, \eqref{eq:S_def}, \eqref{eq:proof_CY_bound}, and~\eqref{eq:proof_mu}, we obtain that
	\begin{align*}
		&W(t+1,{\zeta}_{t+1},\tilde{\zeta}_{t+1},p,\tilde{p})\nonumber\\
		&\leq
		\mu(\|\zeta_{t}-\tilde{\zeta}_{t}\|_P^2 + \gamma\|p-\tilde{p}\|_{S_{t}}^2)\nonumber\\
		&\quad + c_\epsilon\|(B(z_t,\tilde{z}_t)+L(z_t,\tilde{z})D(z_t,\tilde{z}_t))(w_t-\tilde{w}_t)\|_P^2 \nonumber\\
		&\quad + 3\gamma \|D(z_t,\tilde{z}_t)(w_t-\tilde{w}_t)\|^2\nonumber\\
		&\quad + c_\epsilon \|L(z_t,\tilde{z}_t)(y_t-\tilde{y}_t)\|_P^2  + 3\gamma \|y_t-\tilde{y}_t\|^2
	\end{align*}
	for all $t\in\mathbb{I}_{[0,T-1]}$, where $c_\epsilon:={2(1+\epsilon)/\epsilon}$. Due to satisfaction of Assumptions~\ref{ass:bounds_MVT} and \ref{ass:obs}, we can find $Q,R\succ0$ such that
	\begin{align}
		\|\bar{w}\|^2_Q &\geq c_\epsilon\|(B(z,\tilde{z})+L(z,\tilde{z})D(z,\tilde{z}))\bar{w}\|_P^2 \nonumber\\
		&\quad + 3\gamma \|D(z,\tilde{z})\bar{w}\|^2,\label{eq:verify_Q}\\
		\|{\bar{y}}\|_R^2 &\geq c_\epsilon \|L(z,\tilde{z}){\bar{y}}\|_P^2 + 3\gamma \|{\bar{y}}\|^2\label{eq:verify_R}
	\end{align}
	for all $z,\tilde{z}\in\mathbb{Z}$ and all $\bar{w}\in\mathbb{R}^q$, ${\bar{y}}\in\mathbb{R}^r$. Consequently, we can infer that
	\begin{align}
		&W(t+1,{\zeta}_{t+1},\tilde{\zeta}_{t+1},p,\tilde{p})\nonumber\\
		&\leq \mu W(t,{\zeta}_{t},\tilde{\zeta}_{t},p,\tilde{p}) + \|w_{t}-\tilde{w}_{t}\|_{Q}^2 + \|y_{t}-\tilde{y}_{t}\|_{R}^2\label{eq:W_t}
	\end{align}
	for all $t\in\mathbb{I}_{[0,T-1]}$.
	Recursive application of \eqref{eq:W_t} yields
	\begin{align}
		&W(T,{\zeta}_{T},\tilde{\zeta}_{T},p,\tilde{p})\nonumber\\
		& \leq \mu^T W(0,{\zeta}_{0},\tilde{\zeta}_{0},p,\tilde{p})  + \sum_{j=1}^T\mu^{j-1}(\|w_{T-j}-\tilde{w}_{T-j}\|_{Q}^2\nonumber \\
		& \qquad + \|y_{T-j}-\tilde{y}_{T-j}\|_{R}^2).\label{eq:proof_bound_W}
	\end{align}
	Here, $W(T,{\zeta}_{T},\tilde{\zeta}_{T},p,\tilde{p})\geq \gamma\|p-\tilde{p}\|_{S_{T}}^2$ by construction; hence, applying \eqref{eq:proof_St} for $T$ times with $S_0=0$ and using that $\mathcal{C}_T(Z,\tilde{Z})\succeq \alpha I_o$ from \eqref{eq:cond_C} leads to $W(T,{\zeta}_{T},\tilde{\zeta}_{T},p,\tilde{p}) \geq \gamma\alpha\|p-\tilde{p}\|^2$.
	Furthermore, by the definition of $\zeta$ from \eqref{eq:zeta} and the facts that $Y_0=0$ and $S_0=0$, it holds that $W(0,{\zeta}_{0},\tilde{\zeta}_{0},p,\tilde{p}) =  \|x_{0}-\tilde{x}_{0}\|^2_P$.
	In combination, \eqref{eq:proof_bound_W} implies that the trajectories~\eqref{eq:traj} are element of the set $\mathbb{E}_T$ as defined in Definition~\ref{def:obs} for $\eta_{\mathrm{p}}=\mu$, $S_\mathrm{p} = \alpha\gamma I_o$, $Q_\mathrm{p}=Q$, $R_\mathrm{p}=R$, which hence concludes this proof. \qed
\end{pf*}

Verification of~\eqref{eq:cond_C} requires the knowledge of both trajectories in~\eqref{eq:traj}. This is not the case when applied to the estimation problem presented in Section~\ref{sec:MHE_nonPE}, since the true trajectory is generally unknown.
However, we can make local statements based on data from only one of the trajectories that are valid in a surrounding neighborhood.
To this end, we define the closed ball centered at some $Z\in\mathbb{X}\times\mathbb{P}\times\mathbb{U}^T\times\mathbb{W}^T$ of radius $r>0$ by $\mathcal{B}(Z,r):=\{\tilde{Z}\in\mathbb{X}\times\mathbb{P}\times\mathbb{U}^T\times\mathbb{W}^T: \|Z-\tilde{Z}\|\leq r\}$.
Then, we can evaluate~\eqref{eq:cond_C} at $(Z,Z)$ and define
\begin{equation}\label{eq:O_T}
	\mathcal{O}_T(Z) := \mathcal{C}_T(Z,Z).
\end{equation}

\begin{prop}\label{prop:O}
	For $Z\in\mathbb{X}\times\mathbb{P}\times\mathbb{U}^T\times\mathbb{W}^T$, $T\in\mathbb{I}_{\geq1}$, let
	\begin{equation*}
		M(Z,r) := \max_{\tilde{Z}\in\mathcal{B}(Z,r)} \left\|\frac{\partial\mathcal{C}_T}{\partial(Z,\tilde{Z})}(Z,\tilde{Z})\right\|.
	\end{equation*}
	If $\|\mathcal{O}_T(Z)\| \geq \alpha'$ for some $\alpha'>0$, then
	there exists $r>0$ such that $\|\mathcal{C}_T(Z,\tilde{Z})\|\geq \alpha$ for some $\alpha>0$ for all $\tilde{Z}\in\mathcal{B}(Z,r)$.
\end{prop}

\begin{pf*}{Proof.}
	The proof follows similar lines as the proof of \cite[Lemma 4.14]{Flayac2023}.
	For each $Z\in\mathbb{X}\times\mathbb{P}\times\mathbb{U}^T\times\mathbb{W}^T$ and $r>0$, $M(Z,r)$ exists since every function involved is sufficiently smooth and $\mathcal{B}(Z,r)$ is compact.
	By the mean-value theorem and the definition of $M(Z,r)$, we can infer that $\|\mathcal{O}_T(Z)-\mathcal{C}_T(Z,\tilde{Z})\|\leq M(Z,r)r$.
	From the triangle inequality, we also have that
	\begin{equation*}
		\|\mathcal{O}_T(Z)-\mathcal{C}_T(Z,\tilde{Z})\| \geq \|\mathcal{O}_T(Z)\| - \|\mathcal{C}_T(Z,\tilde{Z})\|.
	\end{equation*}
	Combined, we obtain
	\begin{align*}
		\|\mathcal{C}_T(Z,\tilde{Z})\| &\geq \|\mathcal{O}_T(Z)\| - \|\mathcal{O}_T(Z)-\mathcal{C}_T(Z,\tilde{Z})\|\\
		&\geq \alpha'-M(Z,r)r.
	\end{align*}
	Choosing $r>0$ small enough such that $M(Z,r)r<\alpha'$, we have that $\|\mathcal{C}_T(Z,\tilde{Z})\|\geq\alpha$ with $\alpha=\alpha'-M(Z,r)r>0$, which concludes this proof. \qed
\end{pf*}

To summarize, for the trajectory pair~\eqref{eq:traj} and $Z$ and $\tilde{Z}$ from~\eqref{eq:Z} and \eqref{eq:Z_tilde}, we can make the following conclusion: if $\tilde{Z}\in\mathcal{B}(Z,r)$ with $r$ small enough, then $\mathcal{O}_T(Z)\geq\alpha'>0\Rightarrow \mathcal{C}_T(Z,\tilde{Z})>\alpha>0$ by Proposition~\ref{prop:O}, which implies that the trajectory pair~\eqref{eq:traj} is an element of the set $\mathbb{E}_T$ by Proposition~\ref{prop:C_PE}.

\begin{rem}
	Proposition~\ref{prop:O} provides a local result. Applied to the MHE scheme from Section~\ref{sec:MHE_nonPE}, it requires small disturbances $w$ and a good initial guess of ${x}_0$ and ${p}_0$.
	This is a standard condition for testing observability properties in the presence of general nonlinear systems, cf.,~e.g.,~\cite{Sui2011} and~\cite{Flayac2023}.
	Although the condition on $r$ is not explicitly verifiable in the context of state estimation for general nonlinear systems (due to the fact that $r$ is unknown), checking $\mathcal{O}_T(Z)\geq\alpha'$ for some $\alpha'>0$ yields a reliable heuristic to test in practice if a pair $(Z,\tilde{Z})$ is PE and satisfies~\eqref{eq:cond_C} or not, which is also evident in the simulation example in Section~\ref{sec:example}.
	The construction of $\alpha$ in the proof of Proposition~\ref{prop:O} also shows that larger values of $\alpha'$ should be chosen if the estimates are more uncertain and therefore $r$ is expected to be large.
\end{rem}

\begin{rem}
	Compared to conference version~\cite{Schiller2022a}, the proposed method to verify PE of a pair of trajectories is a major relaxation in the sense that it can be applied for a much more general class of nonlinear systems; its only limitation lies in its local nature. However, note that global results can be recovered by considering the same class of systems as in~\cite{Schiller2022a}, i.e., exhibiting the following properties: (i) the dynamics are affine in $p$ and subject to additive disturbances with a linear output map, i.e., $f(w,u,w,p) = \tilde{f}(x,u) + G(x,u)p + Ew$ and $h(x,u,w,p) = Cx + Fw$;
	(ii) changes in $G(x)$ are directly visible in the output;
	(iii) the matrix $\Phi$ in~\eqref{eq:PHI} is constant by a suitable choice of $L$;
	(iv) the sets $\mathbb{X},\mathbb{P},\mathbb{W}$ are compact.
	Then, modifications of the proof of Proposition~\ref{prop:C_PE} according to proof of~\cite[Th.~10]{Schiller2022a} are possible to construct a mapping $\mathcal{C}_T$ similar to~\eqref{eq:cond_C} that satisfies $\mathcal{C}_T(Z,\tilde{Z}) = \mathcal{C}_T(Z,Z) = \mathcal{O}_T(Z)$; i.e, such that $\mathcal{C}_T(Z,\tilde{Z})$ can be rendered independent of one of its arguments (which would correspond to the unknown true trajectory in the context of state estimation).
\end{rem}